\newcommand{\ourtitle}{Blindspots in Python and Java APIs\\Result in Vulnerable Code}

\documentclass[10pt,journal,letterpaper,nofonttune,compsoc,twoside]{IEEEtran}

\makeatletter

\let\proof\@undefined
\let\endproof\@undefined
\makeatother

\usepackage{listings}
\lstdefinelanguage{cs}
{
  morekeywords={abstract,event,new,struct,as,explicit,null,switch
		base,extern,this,bool,false,operator,throw,
		break,finally,out,true,byte,fixed,override,try,
		case,float,params,typeof,catch,for,private,uint,
		char,foreach,protected,ulong,checked,goto,public,unchecked,
		class,if,readonly,unsafe,const,implicit,ref,ushort,
		continue,in,return,using,decimal,int,sbyte,virtual,
		default,interface,sealed,volatile,delegate,internal,short,void,
		do,is,sizeof,while,double,lock,stackalloc,
		else,long,static,enum,namespace,string,
		where, from, select, group,by, having, into, many,
		last, every,
		or, and, on, var, when,let, zip, combine,
		minute,hour,day,week,year,calendar,count, Delay, Until, TakeUntil, Sample, Skip, Throttle},
	  sensitive=true,
	  morecomment=[l]{//},
	  morecomment=[s]{/*}{*/},
	  morestring=[b]",
}

\newcommand\lt[1]{{\lstinline+#1+}} 

\usepackage{algorithm}% http://ctan.org/pkg/algorithms
\usepackage{algpseudocode}% http://ctan.org/pkg/algorithmicx
\usepackage{balance}
\usepackage{url}
\usepackage{relsize}
\usepackage{comment}
\usepackage{multirow}
\usepackage{booktabs}
\usepackage{array}
\usepackage{xspace}
\usepackage{pslatex}
\usepackage{microtype}
\usepackage{amssymb}
\usepackage{amsmath}
\usepackage{amsthm}
\usepackage{amsfonts}
\usepackage{enumitem}
\usepackage{float}
\usepackage{framed}
\usepackage{wrapfig}
\usepackage{ragged2e}
\usepackage{tcolorbox}

\definecolor{javared}{rgb}{0.6,0,0} % for strings
\definecolor{javagreen}{rgb}{0.25,0.5,0.35} % comments
\definecolor{javapurple}{rgb}{0.5,0,0.35} % keywords
\definecolor{javadocblue}{rgb}{0.25,0.35,0.75} % javadoc

\lstdefinestyle{mystyle}{
keywordstyle=\color{javapurple}\bfseries,
stringstyle=\color{javared},
commentstyle=\color{javagreen},
morecomment=[s][\color{javadocblue}]{/**}{*/},
    numberstyle=\tiny\color{black},
    basicstyle=\linespread{0.8}\ttfamily\scriptsize,
    breakatwhitespace=false,
    breaklines=true,
    captionpos=b,
    keepspaces=true,
    numbers=left,
    numbersep=5pt,
    showspaces=false,
    showstringspaces=false,
    showtabs=false,
    tabsize=2
}
\theoremstyle{plain}

\theoremstyle{definition}

\usepackage[tight,footnotesize]{subfigure}
\usepackage{balance}
\usepackage{graphicx}
\DeclareGraphicsExtensions{.pdf}
\usepackage{pifont}
\usepackage[nocompress]{cite}
\usepackage{cancel}
\usepackage{color,colortbl}

\definecolor{LightGray}{gray}{0.9}
\definecolor{Gray}{gray}{0.8}

\newcounter{RQCounter}

\usepackage{tikz}

\newcounter{todocount}[section]

\usepackage{hyperref}
\hypersetup{%
  pdftitle = {Blindspots in Python and Java APIs Result in Vulnerable Code},
  pdfkeywords = {},
  pdfauthor = {Yuriy Brun, Tian Lin, Jessie Elise Somerville, Elisha Myers, Natalie C. Ebner},
  bookmarksnumbered,
  bookmarksopen=false,
  urlcolor=black,
  linkcolor=black,
  citecolor=black,
  colorlinks=true,
  pdfstartview={FitH},
}
\usepackage{cleveref}

\usepackage{stfloats}
\newcounter{myenum}
  {\end{list}}

\begin{document}

\hyphenation{Wil-co-xon Trp-Auto-Re-pair}

\title{\ourtitle}

\author{
Yuriy Brun, 
Tian Lin, 
Jessie Elise Somerville, 
Elisha Myers,
Natalie C. Ebner
\IEEEcompsocitemizethanks{
\IEEEcompsocthanksitem Y.~Brun is with the College of Information and
Computer Sciences at the University of Massachusetts Amherst, Amherst, Massachusetts
01003-9264. Email: \href{brun@cs.umass.edu}{brun@cs.umass.edu}
\IEEEcompsocthanksitem E.~Myers is with the Charles E.\ Schmidt College of
Medicine at the Florida Atlantic University, Boca Raton, Florida, 33431. 
Email: \href{myerse2020@health.fau.edu}{myerse2020@health.fau.edu}
\IEEEcompsocthanksitem T.~Lin, J.~E.~Somerville, and N.~Ebner are with the Department of
Psychology at the University of Florida, Gainesville, Florida 32611-2250.
Email: \href{lintian0527@ufl.edu,JSomerville@dental.ufl.edu,natalie.ebner@ufl.edu}{lintian0527@ufl.edu, jsomerville@dental.ufl.edu, natalie.ebner@ufl.edu}
}
\thanks{}}

\markboth{Brun \MakeLowercase{\emph{et al.}}: Blindspots in Python and Java APIs Result in Vulnerable Code}%, VOL. XX, XXXX}%
{Brun \MakeLowercase{\emph{et al.}}: Blindspots in Python and Java APIs Result in Vulnerable Code}

\IEEEcompsoctitleabstractindextext{
\begin{abstract}
% 
% \justify
Blindspots in APIs can cause software engineers to introduce vulnerabilities,
but such blindspots are, unfortunately, common. We study the effect APIs with
blindspots have on developers in two languages by replicating an
109-developer, 24-Java-API controlled experiment. Our replication applies to
Python and involves 129 new developers and 22 new APIs. We find that using
APIs with blindspots statistically significantly reduces the developers'
ability to correctly reason about the APIs in both languages, but that the
effect is more pronounced for Python.
Interestingly, for Java, the effect increased with complexity of the code
relying on the API, whereas for Python, the opposite was true. This suggests
that Python developers are less likely to notice potential for
vulnerabilities in complex code than in simple code, whereas Java developers
are more likely to recognize the extra complexity and apply more care, but
are more careless with simple code. Whether the developers considered API
uses to be more difficult, less clear, and less familiar did not have an
effect on their ability to correctly reason about them. Developers with
better long-term memory recall were more likely to correctly reason about
APIs with blindspots, but short-term memory, processing speed, episodic
memory, and memory span had no effect. Surprisingly, professional experience
and expertice did not improve the developers' ability to reason about APIs
with blindspots across both languages, with long-term professionals with many
years of experience making mistakes as often as relative novices. Finally,
personality traits did not significantly affect the Python developers'
ability to reason about APIs with blindspots, but less extraverted and more
open developers were better at reasoning about Java APIs with blindspots.
Overall, our findings suggest that blindspots in APIs are a serious problem
across languages, and that experience and education alone do not overcome
that problem, suggesting that tools are needed to help developers recognize
blindspots in APIs as they write code that uses those APIs.
\end{abstract} 

}

\maketitle

\section{Introduction}
\label{sec:Introduction}

\IEEEPARstart{V}{ulnerabilities} are, unfortunately, ubiquitous in modern
software~\cite{Symantec2017}. For example, in 2006, Mozilla had 300 bugs
reported per day~\cite{Anvik06}, and per \url{bugzilla.mozilla.org}, the
situation has not improved since. In 2013, the global cost of debugging was
\$312 billion~\cite{Britton13}. In 2020, the global cost of poor quality in
legacy systems was \$520 billion, and the cost of operational software
failures \$1.56 trillion~\cite{Krasner20}. The fact that programs ship with
both known and unknown bugs is a well known and accepted
fact~\cite{Liblit05}. Web-based software, in particular, suffers from quality
problems, with 76\% of all websites containing software vulnerabilities; 9\%
of them critical~\cite{Symantec2017}.

It is important to note that most vulnerabilities are not from new causes.
New applications often contain instances of vulnerabilities that have been
known for years: 61\% of all web applications contain one of the
vulnerabilities captured by the OWASP Top~10 2013 vulnerability categories
list~\cite{OWASP2013}, such as information leakage, flawed cryptographic
implementations, and carriage-return-line-feed (CRLF)
injection~\cite{SSS2016}. And 66\% of the vulnerabilities represent
programming practices that fail to avoid the top~25 most dangerous
programming errors~\cite{CWE/SANS}. New instances of existing, well-known
vulnerabilities, such as SQL injections and buffer overflows, are frequently
reported in new software~\cite{sqli2017, nvd-buffer-overflow}.

One way in which vulnerabilities find their way into systems is when
developers use application programming interfaces (APIs) in unsafe or
unintended ways. The problem lies, in part, in that APIs can be
counterintuitive. For example, use of \lt{strcpy()}\,---\,known for
nearly three decades~\cite{morrisworm2003} to lead to a buffer overflow
vulnerability if developers do not check and match sizes of the destination
and source arrays\,---\,can often cause \emph{blindspots} in a developer's
mind. As a developer put it, \emph{``It's not straightforward that misusing
strcpy() can lead to very serious problems. Since it's part of the standard
library, developers will assume it's OK to use. It's not called
unsafe\_strcpy() or anything, so it's not immediately clear that that problem
is there.''}~\cite{Oliveira2014}. Using APIs can be difficult. Mapping
requirements to proper API usage protocols, understanding API side effects,
and even deciding between differing expert opinions on API use all pose
challenges~\cite{Gopstein2017, Robillard2009, Robillard2011}. Developers
misunderstanding APIs is frequently the cause of security
vulnerabilities~\cite{Cappos2014, Dagenais2010, rahman2016}.

The question arises of why, if certain APIs are known to cause
vulnerabilities, do developers continue to misuse them. Developers often
blindly trust APIs, which can lead to blindspots, misconceptions,
misunderstandings, or oversights in how APIs are expected to be
used~\cite{Cappos2014}. Developers can even feel that they're outsourcing the
responsibility for ensuring security when using APIs, not feeling themselves
responsible for ensuring the API does the right thing with respect to
security~\cite{Oliveira2014}. These blindspots can lead to violations of the
recommended API usage protocols, and to the introduction of security
vulnerabilities, if, for example, API functions invocations have security
implications that are not readily apparent to the developer.

To study how blindspots affect developers, we recently designed and executed
a controlled study with 109~developers working on programming tasks (called
puzzles) in Java~\cite{Oliveira18soups}. Unlike many studies, more than 70\%
of our participants were professional developers (and less than 30\%
students); the average developer had more than 6 years of programming
experience. We found that (1)~the presence of APIs with known blindspots
reduced developers' accuracy in answering security questions and their
ability to identify potential security concerns in the code, (2)~more complex
code puzzles, as measured by cyclomatic complexity, led to more developer
confusion, (3)~surprisingly, developers' cognitive function and expertise and
experience did not predict their ability to detect API blindspots, but
(4)~developers exhibiting greater openness and lower extraversion were more
likely to detect API blindspots. These findings have allowed research into
understanding why developers make security mistakes~\cite{Parker20,
Votipka20}, gaining insight into the developers' rationale in making API-use
decisions~\cite{Linden20}, and evaluating the usability of security
APIs~\cite{Wijayarathna19}, including as they apply to smart
contracts~\cite{Wan21}. Unfortunately, our prior study had several
limitations. Most notably, it was limited to Java development. As such, while
the findings shed some light on challenges that arise in using APIs, the
study could not determine which findings generalize across languages and
which are specific to Java, or, in fact, only to the API blindspots tested.

To address this shortcoming, in this paper, we replicate that study with 129
new developers, and 22 new puzzles incorporating all new APIs, 16 of which
contain known blindspots. Most importantly, our replication study is entirely
in Python, allowing us to identify observations that are language-specific
versus those that generalize to broader development practices. Our study's
129 developers include professional developers and senior undergraduate and
graduate students (91~professionals, 38~students, 28.6~years old on average,
89.9\%~male). We designed each puzzle to contain a short code snippet
simulating a real-world programming scenario. Of these puzzles, 16 contained
one API function known to cause developers to experience blindspots; we
developed these puzzles based on API functions commonly reported in
vulnerability databases~\cite{nvd, securityfocus} or frequently discussed in
developer fora~\cite{sof}. The other 6 puzzles involved an innocuous API
function. The API functions could each be classified into three categories,
input/output (I/O), cryptography, and string manipulation. Following the
completion of each puzzle, developers responded to one open-ended question
about the functionality of the code and one multiple-choice question that
captured developers' understanding of (or lack thereof) the security
implication of using the specific API function. After completing all puzzles,
each developer provided demographic information and reported their experience
and skills levels in programming languages and technical concepts. Developers
then indicated endorsement of personality statements based on the Five Factor
Personality Traits model~\cite{costa1992} and completed a set of cognitive
tasks from the NIH Cognition Toolbox~\cite{gershon2013nih} and the Brief Test
of Adult Cognition by Telephone (BTACT; modified auditory version for remote
use)~\cite{tun2006}.

Our study, together with the data from our prior, Java-based
study~\cite{Oliveira18soups}, supports the following conclusions:

\begin{itemize}

  \item Developers are statistically significantly less likely to solve
  puzzles with APIs with blindspots than those without blindspots. The effect
  is more pronounced for Python than Java.

  \item The complexity of the puzzle had a different effect on the Python and
  Java developers. For Python, developers were far more often correct when
  solving low-complexity puzzles with APIs without blindspots than with
  blindspots, but there was virtually no difference for high-complexity
  puzzles. For Java, the opposite was true, with developers far more often
  correct when solving high-complexity puzzles with APIs without blindspots
  than with blindspots.

  \item Whether the developers considered the puzzles more difficult, less
  clear, and less familiar did not affect their ability to solve the puzzles.
  
  \item Developers with better long-term memory recall were more likely to
  correctly solve puzzles with blindspots, but short-term memory, processing
  speed, episodic memory, and memory span had no effect.
  
  \item Surprisingly, professional experience and expertice did not improve
  the developers' ability to solve puzzles with APIs with blindspots across
  both languages.
  
  \item Developers with lower extraversion and higher openness as personality
  traits were more accurate in solving Java puzzles with APIs with
  blindspots, but for Python, developers' personality traits were not
  associated with their ability to solve such puzzles.

\end{itemize}

We make public all Java and Python puzzles for use in future research as well
as the surveys, data, and analysis code here:
\url{https://osf.io/ahpfv/?view_only=37978cb1e67941d5b1e572f2fba982c9/}.

The remainder of this paper is organized as follows. 
Section~\ref{sec:API Puzzles} describes the API puzzles.
Section~\ref{sec:methods} presents our study's methodology.
Section~\ref{sec:results} presents our experimental design and assesses the
results, and
Section~\ref{sec:discussion} discusses some of the implications of these
findings.
Section~\ref{sec:related} places this study in the context of related work,
and Section~\ref{sec:contributions} summarizes our contributions.

\section{API Puzzles}
\label{sec:API Puzzles}

Our study is focused around puzzles, snippets of code that simulate a
real-world programming scenario. The snippets of code use APIs, some of which
are known to contain blindspots. 

The goal of a puzzle is to simulate a clear, concise, and unambiguous
programming task representative of a real-world programming task. Users
solving puzzles would necessarily interact with the contained APIs. As some
of the APIs contain known blindspots, while others do not, we aim to use the
users' attempts to solve the puzzles to understand the impact of blindspots
on their behavior. Our study counterbalanced puzzle selection so that so that
each participant received 6 randomly selected puzzles out of a pool of 22, with
four of those puzzles containing a blindspot (details described in
Section~\ref{sec:methods}).

Figure~\ref{fig:sample python puzzle} shows a sample Python puzzle. This code
in the puzzle uses the \lt{os.path.exists} API, which has a known
blindspot. The fact that \lt{print} on lines~3,~6,~and~8 are statements
(not functions), implies that this code is written in Python version 2.x. The
\lt{input} function executes the input typed by the user. If the input
the user types is a Python expression, Python will execute that expression.
This means that the user can type arbitrary code and the program will execute
it, potentially corrupting data or giving up control of the machine\,---\,a
serious, known vulnerability called a \emph{code injection}
attack.\!\footnote{\url{https://www.cvedetails.com/cve/CVE-2018-1000802/}}

\begin{figure}

\begin{tcolorbox}[standard jigsaw,opacityback=0,
colframe=red!75!black]

\textbf{Scenario:}

Consider the following snippet of Python code that searches for a particular
file in a file system, for which a path is entered by the user. The code uses
the \lt{os.path.exists(path)} function, which returns \lt{true} if
the string passed as the parameter refers to an existing path, and
\lt{false} otherwise. Consider the snippet of code below and answer the
following questions, assuming that the code has all required permissions to
execute.

\begin{lstlisting}[language=Python]
import os
print "For which pathname do you want to search?"
# The user enters a file pathname to search
path = input('> ')
if os.path.exists(str(path)):
  print path, "pathname exists."
else:
  print path, "pathname does not exist."
\end{lstlisting}

\textbf{Questions}:
\begin{enumerate}[leftmargin=1.4em]
\item What will the program do when executed?

\item What will happen if the user enters the string
``\lt{os.system("date")}'' when prompted for the file pathname?

\begin{enumerate}
\item The current date will be displayed on the terminal.
\item The program will print ``pathname does not exist.''
\item The program will crash with error message ``invalid input''.
\item The program will crash with no error message.
\item None of the above.
\end{enumerate}
\end{enumerate}
\end{tcolorbox}

\caption{A sample Python puzzle. The \lt{input} API contains a known
blindspot: the user's unfiltered input can be executed directly by the
program. Because the \lt{os.system} function returns 0, (though this
depends on the operating system), the correct answer to question~1 is that the
program prints ``0 pathname exists.'' if there is a file named ``0'' in the
current directory, and ``0 pathname does not exist.'' otherwise. The correct
answer to question 2 is ``a''.}

\label{fig:sample python puzzle}
\end{figure}

Each puzzle, as the one in Figure~\ref{fig:sample python puzzle}, consists of
three elements: the scenario description, the code, and two questions.

We collected API functions commonly reported in vulnerability
databases~\cite{nvd, securityfocus} or frequently discussed in developer
fora~\cite{sof}, and used these APIs to create puzzles, small programs that
exercise the APIs. We created a total of 22 Python puzzles; 16 contained APIs
with blindspots and 6 contained APIs without blindspots using innocuous API
functions not known to cause vulnerabilities. Figure~\ref{fig:sample python
puzzle no blindspot} shows an example puzzle that uses an API without a known
blindspot.

\begin{figure}

\begin{tcolorbox}[standard jigsaw,opacityback=0,
colframe=red!75!black]

\textbf{Scenario:}

Consider this snippet of Python code that implements a music directory by
storing information about song titles, artist names, album names, and release
dates in a MySQL database. The user is asked to enter the genre of music for
which they are searching. This input is then passed to a database, which uses
this information to select songs and returns them to the user. Consider the
snippet of code below and answer the following questions, assuming that the
code has all required permissions to execute.

\begin{lstlisting}[language=Python]
# Import whatever is needed
genre = input("Please enter your desired genre>")
# Open database connection
db = MySQLdb.connect("localhost","allUsers","54373$!@","MUSICDB")
'''
Prepares a cursor object that will hold the 
information on the rows accessed by our query
'''
cursor = db.cursor()
cursor.execute("SELECT songName, albumName, releaseDate "
    "FROM Songs WHERE genre = '%s'", genre)
# Get all the rows
results = cursor.fetchall()
for row in results:
   print(row)
db.close()
\end{lstlisting}

\textbf{Questions}:
\begin{enumerate}[leftmargin=1.4em]
\item What will the program do when executed?

\item What type of information could a user retrieve from this program? 

\begin{enumerate}
\item The song name
\item The album name
\item The artist name
\item A and B
\item Any attribute from the database
\end{enumerate}
\end{enumerate}
\end{tcolorbox}

\caption{A sample Python puzzle that uses the \lt{.execute} API that does
not contain a known blindspot. The correct answer to question~1 is that the
function will return records from the database that match the specified
genre. The correct answer to question~2 is ``d''.}

\label{fig:sample python puzzle no blindspot}
\end{figure}

The APIs used in the puzzles come from three categories, input/output (I/O),
cryptography, and string manipulation. I/O APIs involve operations such as
networking activity, and reading and writing from and to streams, files, and
internal memory buffers. Cryptography APIs include encryption, decryption,
and key agreement. String manipulation APIs include editing and processing
strings, such as queries and user input.  

We designed the puzzles aiming to keep each puzzle's complexity to the
minimum necessary to properly capture the API's use. As a result, the
complexity of the puzzles varied. We measured the puzzles complexity using
cyclomatic complexity, a quantitative measure of the number of linearly
independent paths in the source code~\cite{McCabe1976}. We classified the
puzzles as \emph{low} complexity (cyclomatic complexity of 1--2),
\emph{medium} complexity~(3--4), and high complexity~($\geq5$). Of the 22
puzzles, 8 were low complexity, 11 medium complexity, and 3 high complexity.

Our earlier, Java-based study~\cite{Oliveira18soups} used a total of 24
puzzles, written in Java (16 of which used APIs with blindspots and 8 uses
innocuous APIs). Figure~\ref{fig:sample java puzzle} shows a sample Java
puzzle with an API with a blindspot. This puzzle uses the \lt{Runtime.exec}
API, which executes code passed to it via its argument, also exposing the
\lt{setDate} method to a code injection attack.

\begin{figure}[t]
\begin{tcolorbox}[standard jigsaw,opacityback=0,
colframe=red!75!black]

\textbf{Scenario:}

You are asked to review a utility method written for a web application. The
method, \lt{setDate}, changes the date of the server. It takes a
\lt{String} as the new date (``dd-mm-yyyy'' format), attempts to change
the date of the server, and returns \lt{true} if it succeeded, and
\lt{false} otherwise. Consider the snippet of code below (assuming the
code runs on a Windows operating system) and answer the following questions,
assuming that the code has all required permissions to execute.

\begin{lstlisting}[language=Java]
// OMITTED: Import whatever is needed
public final class SystemUtils {
  public static boolean setDate (String date)
      throws Exception {
    return run("DATE " + date);
  }

  private static boolean run (String cmd)
      throws Exception {
    Process process = Runtime.getRuntime().exec("CMD /C " + cmd);
    int exit = process.waitFor();

    if (exit == 0)
      return true;
    else
      return false;
  }
}
\end{lstlisting}

\textbf{Questions}:
\begin{enumerate}[leftmargin=1.4em]
\item What will the \lt{setDate} method do when executed?

\item If a program calls the \lt{setDate} method with an arbitrary
\lt{String} value as the new date, which one statement is correct?

\begin{enumerate}
\item If the given \lt{String} value does not conform to the
      ``dd-mm-yyyy'' format, an exception is thrown.
\item The \lt{setDate} method cannot change the date.
\item The method might do more than changing the date.
\item The return value of the \lt{waitFor} method is not interpreted
      correctly (lines 14~17).
\item The web application will crash.
\end{enumerate}
\end{enumerate}
\end{tcolorbox}

\caption{A sample Java puzzle. This puzzle uses the \lt{Runtime.exec}
API, which executes code passed to it via its argument. Because the
\lt{run} method, called by the \lt{setDate} method simply passes
\lt{setDate}'s arbitrary \lt{String} argument to
\lt{Runtime.exec}, this method may end up executing arbitrary code. The
correct answer to question 2 is ``c''.}

\label{fig:sample java puzzle}
\end{figure}

All Java and Python puzzles, as well as the surveys, data, and analysis code,
are available here:
\url{https://osf.io/ahpfv/?view_only=37978cb1e67941d5b1e572f2fba982c9/}.

\section{Data-Collection Methodology}
\label{sec:methods}

We first describe our study's participants (Section~\ref{sec:Participants}),
and then our data-collection procedure (Section~\ref{sec:Procedure}).

\subsection{Participants}
\label{sec:Participants}

Our study targeted developers who actively use the Python programming
language. We recruited 129 experienced Python developers. We considered
individuals with more than one year of experience in a professional setting
to be \emph{professionals}; we considered all others to be \emph{students}.
Collectively, we refer to all participants as \emph{developers} in this
study. Recruitment methods included flyers and handouts, social media
advertisements, advertisements through companies and their human resources
departments, university listservs, and contacts via the authors' personal
networks, as well as word of mouth. The protocol was approved by the
University of Florida Institutional Review Board. Professionals were
compensated \$50 and students \$20 for study completion. These different
compensation amounts were based on level of programming experience, and were
approved by the Institutional Review Board; they were justified given a
larger financial incentive necessary to recruit professional developers, in
consideration of their relatively high-paying jobs and a more limited
availability.

Initially, we received a total of 423 emails from interested developers. Out
of the 423, 13~(3.1\%) participants explicitly dropped out and 184~(43.5\%)
participants implicitly dropped out by becoming unresponsive at some point in
the data-collection process. The remaining 226 developers~(53.4\%) completed
the study. Of those 226 developers, 84~(37.2\%) developers' data was
incomplete, e.g., they did not answer some of their puzzles' questions, did
not complete the audio task for cognitive assessment (described in
Section~\ref{sec:RQ4and5and6}), or had technical difficulties, such as
browser incompatibility issues, that prevented audio recording. We discarded
these 84~developers' data. During the enrollment process, we discovered that
13~(5.6\%) of the developers were repeat participants; that is, they
participated multiple times despite instructions not to do so. For those
individuals, we accepted their first data submission as a valid entry and
discarded all subsequent submissions. That resulted in 129 valid developer
sessions. Unless otherwise stated, we report our results based on a that
sample of 129~developers who proceeded through all study procedures as
instructed and completed all tasks.

Figure~\ref{fig:participants} summarizes the participant demographics and
their professional expertise and experience. The 129~developers consisted of
91~(70.5\%) professional developers and 38~(29.5\%) 
students. The vast majority of developers (125,~96.9\%) had
2 or more years of Python programming experience. 
Student participants
self-reported a relatively high programming experience (mean of 5.3 years,
standard deviation of 2.9 years), likely a result of programming prior to
entering the university or being students for more than six years (e.g., PhD
students). Participants ranged between the ages of 18 and 71 years (mean of
28.4 years, standard deviation of 7.8 years), with the large majority of
participants being male (116,~89.9\%). Participants were recruited across the
globe with concentrations in North and South America (74,~57.4\%), Asia
(43,~33.3\%), and Europe (11,~8.7\%). Overall, participants came from the
United States, Brazil, India, Bangladesh, Pakistan, Greece, Poland, France,
England, Germany, etc.

\begin{figure}

\resizebox{\columnwidth}{!}{
\begin{tabular}{lll}
\toprule
                                       & \textbf{Professionals}      & \textbf{Students}    \\
\textbf{Total}                         & 91\phantom{ (00.0\%)}  & 38\phantom{ (00.0\%)}\\
\textbf{Years of Programming}          & \phantom{0}6.9  (stdev 5.6)      & \phantom{0}4.1 (stdev 2.1)  \\ 
\textbf{Age}                           & 30.1 (stdev 7.2)      & 25.1 (stdev 4.4) \\
\multicolumn{3}{l}{\textbf{Gender}} \\
~~Male (116)                           &  82 (90.1\%)          & 34 (89.5\%)      \\
~~Female (13)                          &  \phantom{0}9 \phantom{0}(9.9\%)            & \phantom{0}4 (10.5\%)       \\
\multicolumn{3}{l}{\textbf{Highest Degree Earned}} \\
~~High School or GED                     &  \phantom{0}2 \phantom{0}(2.2\%)  & \phantom{0}3 \phantom{0}(7.9\%)        \\
~~Some College                           &  \phantom{0}1 \phantom{0}(1.1\%)  & \phantom{0}3 \phantom{0}(7.9\%)        \\
~~Associates / 2-year degree             &  \phantom{0}1 \phantom{0}(1.1\%)  & \phantom{0}1 \phantom{0}(2.6\%)        \\
~~Bachelor's / 4-year degree             &  41 (45.1\%)                      & 16 (42.1\%)      \\
~~Some Graduate School                   &  \phantom{0}5 \phantom{0}(5.5\%)  & \phantom{0}5 (13.1\%)       \\
~~Graduate-Level Degree                  &  41 (45.1\%)                      & 10 (26.3\%)    \\
\multicolumn{3}{l}{\textbf{Annual Income}} \\
~~\phantom{00,000}\$0--\phantom{0}\$39,999 &  43 (47.2\%)          & 32 (86.8\%)    \\
~~\phantom{0}\$40,000--\phantom{0}\$70,000 &  20 (22.0\%)          & \phantom{0}4 (10.8\%)     \\
~~\phantom{0}\$70,000--\$100,000           &  15 (16.5\%)          & \phantom{0}0 \phantom{0}(0.0\%)      \\
~~\$100,001--\$200,000                     &  11 (12.1\%)          & \phantom{0}0 \phantom{0}(0.0\%)      \\
~~\$201,000+                               &  \phantom{0}2 \phantom{0}(2.2\%)            & \phantom{0}1 \phantom{0}(2.6\%)      \\
\multicolumn{3}{l}{\textbf{Ethnicity}} \\
~~American Indian or Alaskan             &  \phantom{0}1 \phantom{0}(1.1\%) & \phantom{0}0 \phantom{0}(0.0\%)        \\
~~Asian                                  &  58 (63.7\%)                     & 23 (60.5\%)      \\
~~Black or African American              &  \phantom{0}0 \phantom{0}(0.0\%) & \phantom{0}0 \phantom{0}(0.0\%)        \\
~~Hawaiian or other Pacific Islander     &  \phantom{0}0 \phantom{0}(0.0\%) & \phantom{0}0 \phantom{0}(0.0\%)        \\
~~White                                  &  24 (26.4\%)                     & 12 (34.2\%)      \\
~~Other or multi-racial                  &  \phantom{0}8 \phantom{0}(8.8\%) & \phantom{0}2 \phantom{0}(5.3\%)        \\
\multicolumn{3}{l}{\textbf{Location}} \\
~~North \& South America                 &  46 (50.5\%)                       & 28 (73.7\%)      \\
~~Asia                                   &  38 (41.8\%)                       & \phantom{0}6 (15.8\%)       \\
~~Europe                                 &  \phantom{0}7 \phantom{0}(7.7\%)   & \phantom{0}4 (10.5\%)       \\
\bottomrule

\end{tabular}
}

\caption{Demographics and professional expertise and experience of the study
participants.}
\label{fig:participants}
\end{figure}

\subsection{Data Collection Procedure}
\label{sec:Procedure}

Data collection started in August 2017 and ended in August 2018. After
initial contact with interested developers, we used an online screening
questionnaire to determine study eligibility and compensation (e.g.,
sufficient knowledge of Python, fluency in the English language, age over
18 years).
Exclusion criteria included previous participation in a similar study using
Java programming language~\cite{Oliveira18soups} or previous participation in
the current study, no knowledge of Python, under 18 years of age, lack of
proficiency in English, and unwillingness to install and use the latest
version of Mozilla Firefox, as our survey was tested on and compatible with
this browser.

Eligible developers received a digital informed consent form, which disclosed
study procedures, the minimal risk from study participation, and data privacy
and anonymity. Each developer was assigned a unique, anonymized identifier to
assure confidentiality. After providing their digital signature, developers
received an audio test link. This test was implemented to ensure that proper
data quality was captured for the audio recording during the cognitive
testing. It was instituted following the failure to collect audio data from a
large fraction of the participants in the earlier Java
study~\cite{Oliveira18soups}. Participants were instructed to read three
sentences into their microphone. The resulting test audio was reviewed by
research staff, and if the audio quality was good, participants received a
personalized link to the online assessment. If the audio was not captured or
of poor quality, research staff worked with the participant to address audio
issues. This revision of the data collection methodology was successful in
addressing the shortcomings of the earlier study~\cite{Oliveira18soups}.

After participants completed the main survey, they were sent an additional
survey on software security. This survey aimed to gather information
regarding the participants' respective organizations' focus, organization
size, organization software products developed, and also asked
security-related questions.

Developers were strongly encouraged to complete the study in two separate
sittings to counteract possible fatigue effects (one sitting to work on the
puzzles and complete the demographic questionnaire, and the other sitting to
complete the psychological and cognitive assessment). All data collection
took place in a location of the participants' choosing. To increase
ecological validity, participants were informed they could use outside
resources as assistance and were asked to report the resources used
throughout the survey. However, they were not informed that the study was
about code security.

The study procedure comprised of five parts. The first part, programming
puzzles (recall Section~\ref{sec:API Puzzles}) involved responding to
the programming puzzles and related questions. Participants were told that
the puzzles were designed to examine how developers interpret and reason
about code; they were not informed about the nature or presence of blindspots
in the code. Identical to our prior, Java-based study~\cite{Oliveira18soups},
the second part of the study assessed participant demographics; the third
part professional experience and expertise; the fourth part was a personality
assessment that utilized the Big Five Inventory (BFI)
questionnaire~\cite{srivastava99}; and the final part consisted of a
cognitive assessment, comprising the Oral Symbol Digit Test from the NIH
Toolbox~\cite{gershon2013nih} and the Brief Test of Adult Cognition by
Telephone (BTACT)~\cite{tun2006}. The average time to complete the entire
study ranged between 30 to 90 minutes.

\section{Experimental Design and Results}
\label{sec:results}

Our analysis aims to answer six research questions:

\begin{tcolorbox}
RQ1: Are developers less likely to correctly solve puzzles with API functions
containing blindspots than puzzles with innocuous functions? Does the
underlying programming language have an effect?

\smallskip
RQ2: Do developers perceive puzzles with API functions containing blindspots
as more difficult, as less clear, and as less familiar than non-blindspot
puzzles? Does the underlying programming language have an effect?

\end{tcolorbox}
\begin{tcolorbox}
RQ3: Are developers less confident about their puzzle solution when working
on puzzles with API functions containing blindspots than non-blindspot
puzzles? Does the underlying programming language have an effect?

\smallskip
RQ4: Are developers with higher cognitive functions (e.g., reasoning, working
memory, and processing speed) better at solving puzzles with API functions
containing blindspots? Does the underlying programming language have an
effect?

\smallskip
RQ5: Are developers with more professional experience and expertise better at
solving puzzles with API functions containing blindspots? Does the underlying
programming language have an effect?

\smallskip
RQ6: Are developers with higher levels of conscientiousness and openness and
lower levels of neuroticism, extraversion, and agreeableness better at
solving puzzles with API functions containing blindspots? Does the underlying
programming language have an effect?
\end{tcolorbox}

\subsection{Analysis Methodology}
\label{sec:Analysis Methodology}

In answering RQ1, RQ2, and RQ3, we used multilevel modeling. In answering
RQ4, RQ5, and RQ6, we used ordinal logistic regression. We conducted all
analyses using Stata 15.0 and applied the Wald test to determine statistical
significance of main and interaction effects. 

Unlike our prior, Java-based study~\cite{Oliveira18soups}, for all
non-significant effects, we further conducted Bayesian statistic analyses to
determine whether a given non-significant effect was more likely due to
insensitivity of the data or reflected a preference to accept the null
hypothesis~\cite{Dienes14}. In other words, for situations in which the
$p$-value of the Wald test was too high for us to reject the null hypothesis,
this analysis allowed us to estimate the likelihood that the $p$-value is high
because of insufficient data versus because there is no underlying difference
between the distributions being compared. A non-significant effect with a
Bayes Factor between 0.33 and 1 indicates data insensitivity (that is, our
data is insufficient to draw a conclusion); a non-significant effect with a
Bayes Factor lower than 0.33 indicates preference to accept the null
hypothesis (that is, that the two distributions being compared actually come
from the same underlying distribution). As a hypothetical example, consider
the situation in which we have two groups of developers participating in our
study, one wearing red hats and one wearing green hats. The Wald test
determines whether the difference in the number of puzzles the two groups
solve correctly is statistically significant\,---\,that is, the $p$-value is
the probability that the two distributions actually come from the same
underlying distribution of developers. If the $p$-value is sufficiently low, we
can conclude that there exists a statistically significant difference between
red-hat and green-hat developers, with respect to the number of puzzles
solved correctly. But if the $p$-value is too high, the Bayes statistic
analysis allows us to determine the likelihood that the two distributions of
developers are actually coming from the same underlying distribution, versus
the possibility that our data is simply insufficient to tell if the
distributions are, in fact, different. This extension of the methodology used
in the original Java-based study we are replicating allows for a more
thorough analysis of our results.  

We now answer each of the six research questions. For each question, we
present the data and analysis for the Python puzzles, recap and compare to
the findings with respect to the Java puzzles~\cite{Oliveira18soups}, and
analyze the languages' effect. For the direct comparison between the two
programming languages, we applied the same analytic approach as just
described, but added programming language (Python versus Java) as a moderator
in the models. For these analyses we were particularly interested in the
interaction of programming language in each model.

\subsection{Do Blindspots Make Programming Tasks More Difficult?}
\label{sec:RQ1}

To answer whether developers are less likely to correctly solve puzzles with
API functions containing blindspots than puzzles without blindspots (RQ1), we used
multilevel logistic regression.  The underlying data were hierarchical:
each set of six puzzles (level 1 data), nested within each developer (level 2
data). The independent variable was the presence of a blindspot (0 for no
blindspot, 1 for blindspot), and the dependent variable was whether the
developer solved the puzzle (0 for incorrect, 1 for correct). In this model,
we also considered the random effect of the intercept to accommodate for
inter-individual differences in overall puzzle accuracy. 

For Python, the 129 participants correctly solved 74\% of the puzzles without
blindspots, but only 36\% of the puzzles with blindspots. The effect of the
presence of blindspot was significant (Wald $\chi^2(1) = 80.61$, $p <
0.001$), rejecting the null hypothesis that the puzzle solving accuracy for
puzzles with blindspots and without came from the same distribution. Thus,
participants were more than twice as likely to solve a Python puzzle
correctly if that puzzle used an API without a blindspot than if it used one
with a blindspot.

\begin{figure}[t]

\includegraphics[width=\columnwidth]{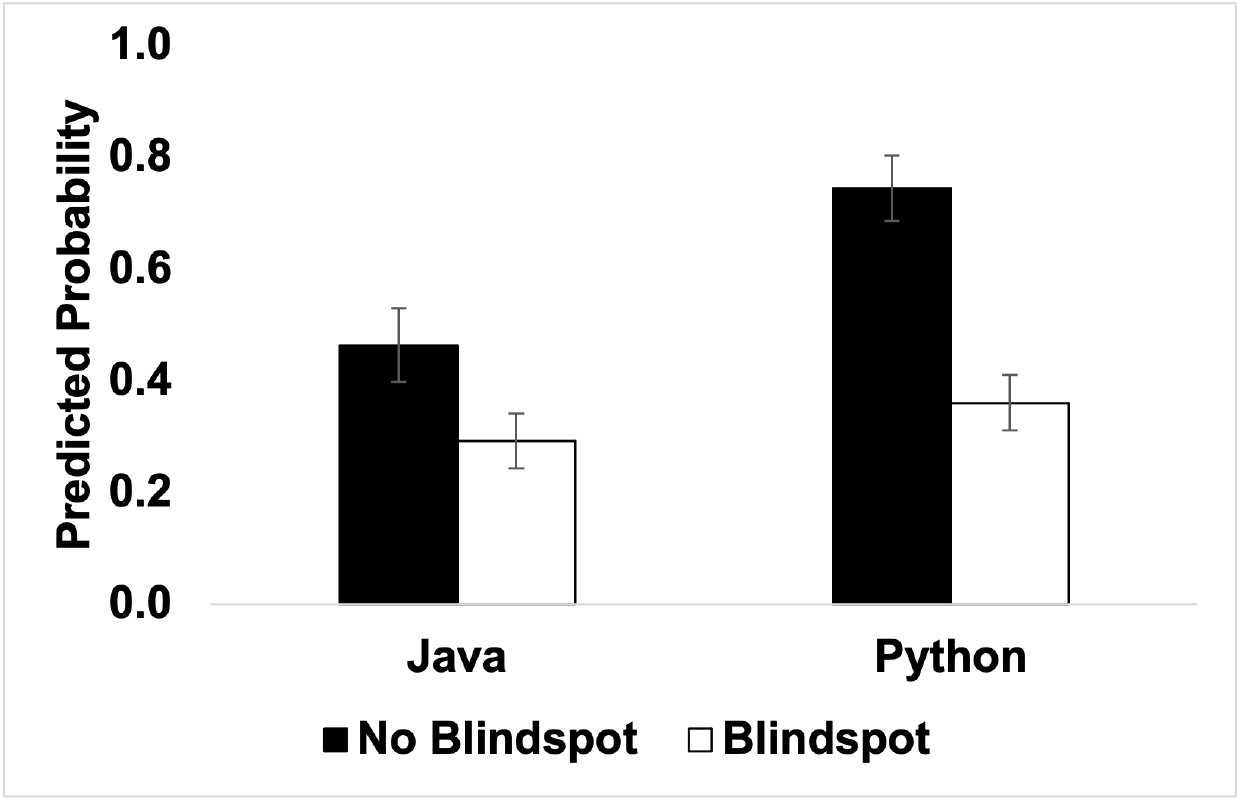}

\caption{Interaction effect of the programming language and the presence of a
blindspot on puzzle accuracy. The x-axis shows the two programming languages,
Java and Python. The y-axis shows the predicted accuracy (predicted
probability of correctly solving a puzzle). Error bars represent 95\%
conﬁdence intervals.}
\label{fig:langAccuracy} 
\end{figure}

These results are similar to the observations for Java API
puzzles~\cite{Oliveira18soups}: the 109 participants correctly solved 46\% of
the puzzles without blindspots, but only 29\% of the puzzles with blindspots,
which was similarly a significant difference (Wald $\chi^2(1) = 20.60$, $p <
0.001$). Thus, participants were about 1.6 times as likely to solve a Java
puzzle correctly if that puzzle used an API without a blindspot than if it
used one with a blindspot.

We combined the Java and Python datasets and treated the programming language
(Python versus Java) as a moderator in the models. Expectedly, we again found
a statistically significant difference in puzzle accuracy for puzzles with
and without APIs with blindspots (Wald $\chi^2(1) = 94.84$, $p < 0.001$). The
interaction between programming language and presence of blindspot (Wald
$\chi^2(1) = 13.93$, $p < 0.001$) was significant.
Figure~\ref{fig:langAccuracy} depicts this interaction, and shows that,
overall, developers were less likely to accurately solve puzzles with
blindspots than ones without blindspots, with this effect more pronounced for
Python ($B = -1.78$ (the slope of the line between the predictor variable and
the dependent variable), $z = -8.98$ (z-score describes the deviation from
the mean in number of standard deviations), $p < 0.001$, odds ratio $= 0.17$)
than for Java ($B = -0.81$, $z = -4.54$, $p < 0.001$, odds ratio $= 0.44$).

\begin{tcolorbox}
(RQ1) Overall, our statistical analysis supports that developers were
significantly more successful in correctly solving puzzles that used APIs
without blindspots than with blindspots, for both programming languages, with
this effect more pronounced for Python than Java.
\end{tcolorbox}

We next looked at the three kinds of APIs used in our puzzles (I/O,
cryptography, and string manipulation) and the cyclomatic complexity of the
puzzles, and how these variables affected the developers' ability to solve
puzzles. To control for family-wise type-I error inflation due to testing of
multiple dependent models (i.e., models that share the same dependent
variable), we applied Bonferroni correction for the $p$-value threshold to
determine statistical significance ($p < 0.025$).

\begin{figure}[t]

\includegraphics[width=\columnwidth]{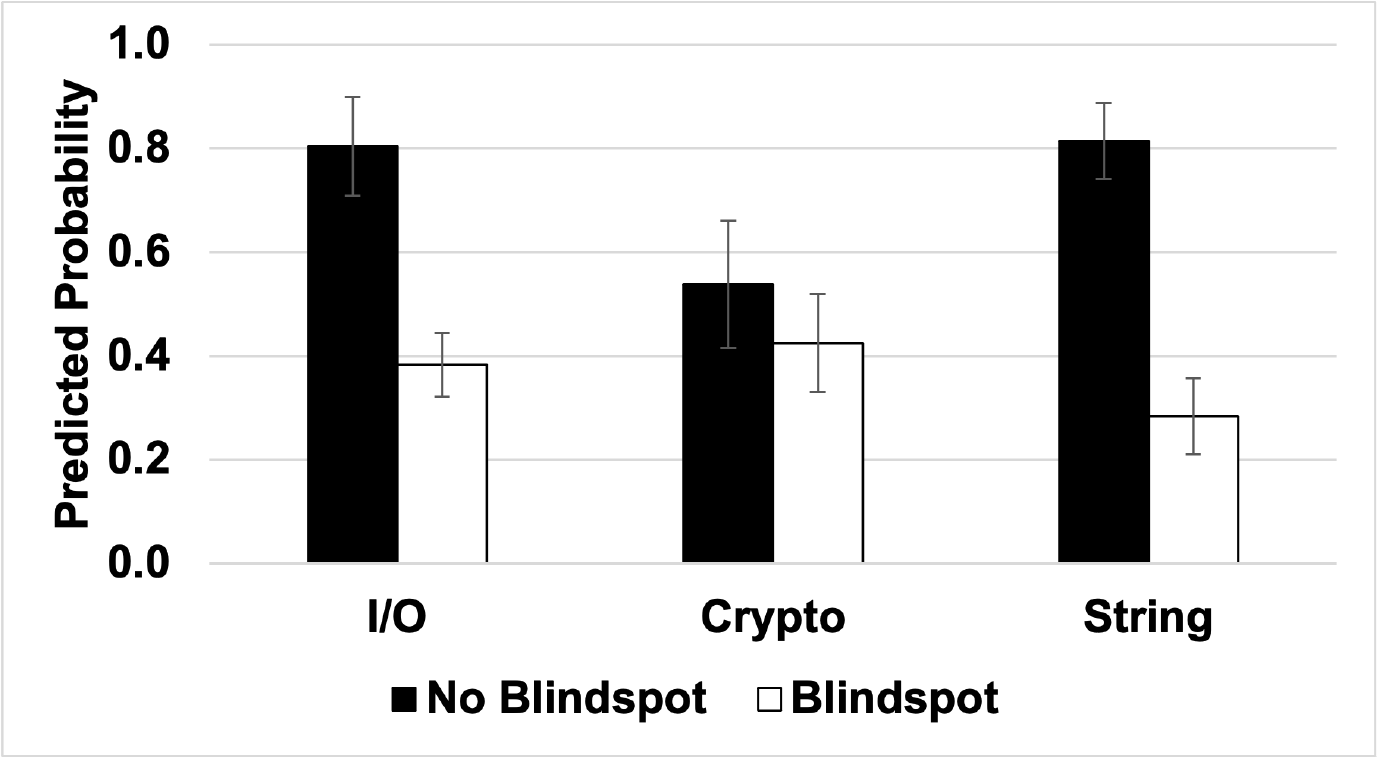}

\caption{Interaction effect of the API type and the presence of a blindspot
on puzzle accuracy. The x-axis shows the three types of API usage: I/O,
crypto, and string manipulation. The y-axis shows the predicted accuracy
(predicted probability of correctly solving a puzzle). Error bars represent
95\% conﬁdence intervals after Bonferroni correction of the $p$-value.}
\label{fig:typeAccuracy} 
\end{figure}

For Python, adding the categorical variable API usage type, we measured its
interaction with the presence of an API with a blindspot in a puzzle as a
predictor for whether the developer correctly solved the puzzle. The main
effect of API usage type was not significant (Wald $\chi^2(2) = 6.55, p =
0.04$). However, its interaction with the presence of a blindspot was
significant (Wald $\chi^2(2) = 21.06, p < 0.001$).
Figure~\ref{fig:typeAccuracy} shows that while developers were always less
likely to accurately solve puzzles with APIs containing blindspots than
puzzles without blindspots, this difference in accuracy was more pronounced
for puzzles using API functions that involved I/O and string manipulation
than those that involved crypto.

In the Java-puzzle study~\cite{Oliveira18soups}, the main effect of the
presence of blindspot was not significant (Wald $\chi^2(2) = 0.91$, $p =
0.34$), but the main effect of API usage type (Wald $\chi^2(2) = 10.64$, $p =
0.005$) and its interaction with the presence of blindspot (Wald $\chi^2(2) =
24.81$, $p < 0.001$) were significant. Accuracy was higher for puzzles
without blindspots than for ones with blindspots with an API function that
involved I/O. Accuracy was comparable in both puzzles with and without
blindspots with API functions that involved Crypto, String manipulation.

\begin{figure}[t]

\includegraphics[width=\columnwidth]{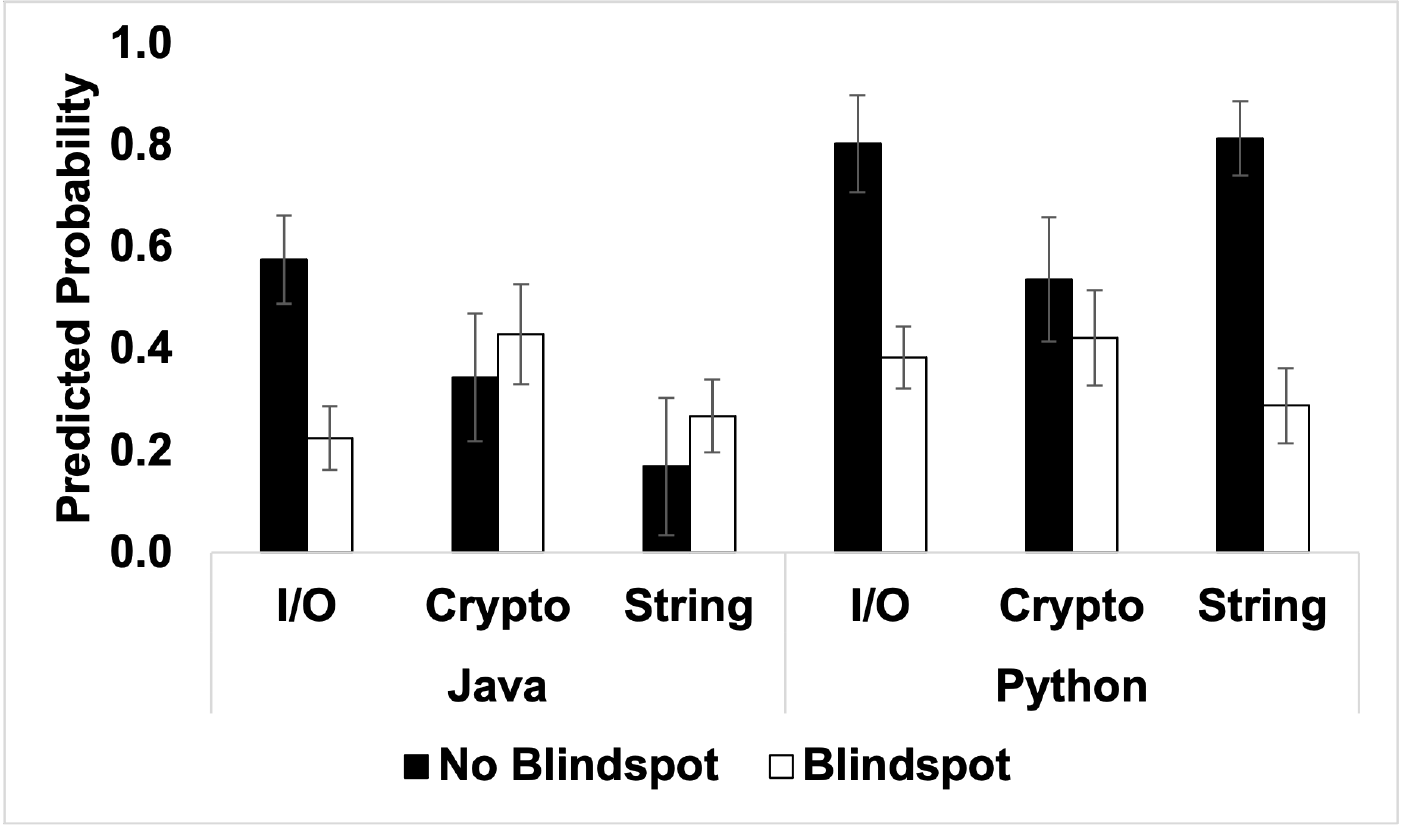}

\caption{Interaction effect of the programming language, API usage type, and
the presence of a blindspot on puzzle accuracy. The x-axis shows the three types of
API usage: I/O, Crypto, and String, and differentiates between Java~(left)
and Python~(right) puzzles. The y-axis shows the predicted accuracy
(predicted probability of correctly solving a puzzle). Error bars represent
95\% conﬁdence intervals after Bonferroni correction of the $p$-value.}
\label{fig:typeJointAccuracy} 
\end{figure}

We then, again, treated the language as a moderator to explore the extent to
which API usage type moderated accuracy for puzzles with blindspots versus
ones without blindspots for Python and Java. Again, we applied Bonferroni
correction for the $p$-value threshold to determine statistical significance
($p < 0.025$). The three-way interaction between presence of blindspot, API
usage type, and programming language was significant (Wald $\chi^2(2) =
12.38$, $p = 0.002$). Figure~\ref{fig:typeJointAccuracy} shows that, while in
Python, developers were less likely to correctly solve puzzles with
blindspots than without blindspots for all three API usage types, in Java
this accuracy difference only held in puzzles with I/O API functions, but not
puzzles with Crypto and String API functions.

\begin{tcolorbox}
(RQ1, for different API types) Our statistical analysis supports that for
Python, developers were overall more accurate solving puzzles without
blindspots than puzzles with blindspots for all three API usage types, but
for Java, that was only the case for puzzles with I/O API functions, not
puzzles with Crypto and String API functions.
\end{tcolorbox}

\begin{figure}[t]

\includegraphics[width=\columnwidth]{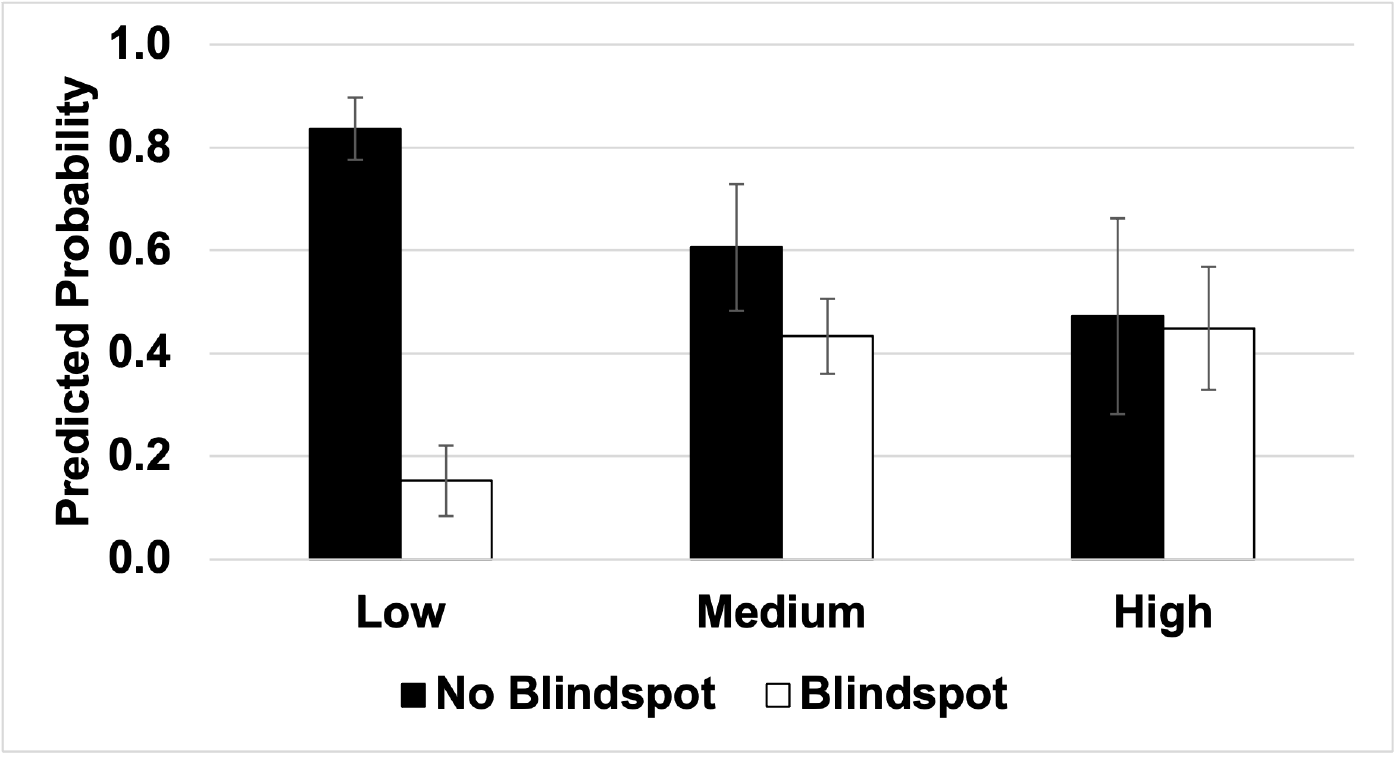}

\caption{Interaction effect of cyclomatic complexity and presence of a
blindspot on puzzle accuracy. The x-axis shows the puzzle's level of
cyclomatic complexity. The y-axis shows the predicted accuracy (predicted
probability of correctly solving a puzzle). Error bars represent 95\%
conﬁdence intervals after Bonferroni correction of the $p$-value.}
\label{fig:complexityAccuracy} 
\end{figure}

Next, adding the categorical variable cyclomatic complexity (1 for low, 2 for
medium, 3 for high), we measured its interaction with the presence of an API
with a blindspot in a puzzle as a predictor for whether the developer
correctly solved the puzzle. Again, while the main effect of cyclomatic
complexity was not significant (Wald $\chi^2(2) = 0.97, p = 0.62$), its
interaction with the presence of a blindspot was significant (Wald $\chi^2(2)
= 37.34, p < 0.001$). Figure~\ref{fig:complexityAccuracy} shows that for
puzzles of low cyclomatic complexity, the difference in accuracy was very
large, whereas for puzzles of medium complexity, the difference in accuracy
was small. For puzzles with high cyclomatic complexity, there was no measured
difference. One possible explanation for this behavior is that developers may
be more confident when dealing with seemingly simpler tasks, and that
confidence may cause them to overlook blindspots more frequently. Meanwhile,
for more complex tasks, the developer may already spend more time and think
more carefully about the code, and thus is more likely to consider the
effects of the APIs with blindspots, falling victim to those blindspots less
frequently. If true, this hypothesis suggests that the effects of at least
some blindspots can be overcome simply with paying extra attention or
thinking through the involved issues more carefully, as opposed to requiring
significant education or even expertise with that particular API.

In the Java-puzzle study~\cite{Oliveira18soups}, the main effect of
cyclomatic complexity was not significant (Wald $\chi^2(2) = 0.74$, $p <
0.69$), but the main effect of the presence of a blindspot (Wald $\chi^2(2) =
23.95$, $p < 0.001$) and its interaction with cyclomatic complexity (Wald
$\chi^2(2) = 30.1$, $p <0.001$) were significant. Accuracy was higher for
puzzles without blindspots than those with blindspots at medium cyclomatic
complexity, and, even more pronounced at high cyclomatic complexity. That is,
the higher the cyclomatic complexity of the code in a puzzle containing APIs
with blindspots, the less likely developers were to correctly solve the
puzzle.

\begin{figure}[t]

\includegraphics[width=\columnwidth]{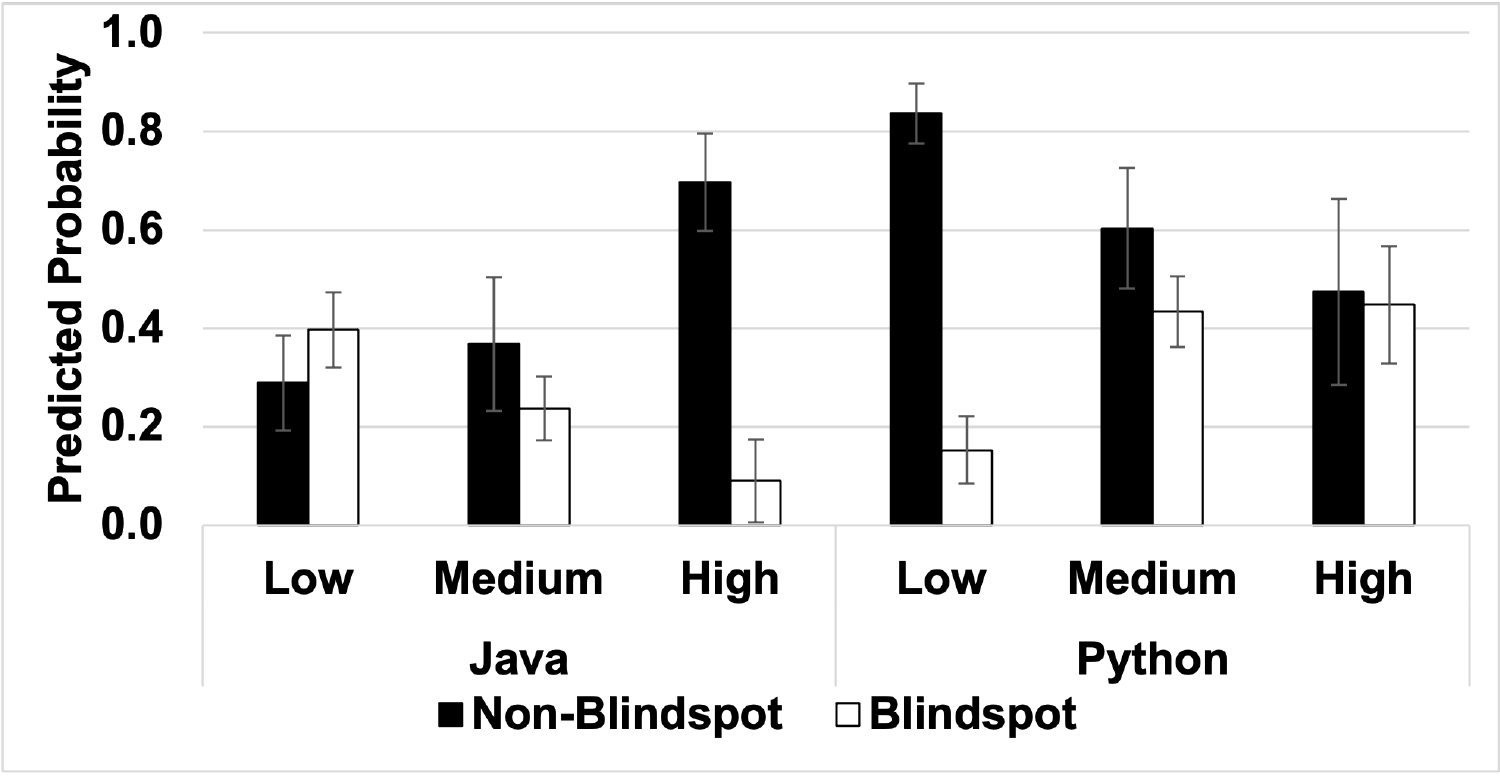}

\caption{Interaction effect of the programming language, cyclomatic
complexity, and presence of a blindspot on puzzle accuracy. The x-axis shows
the puzzle's level of cyclomatic complexity and differentiates between Java
(left) and Python (right) programmers. The y-axis shows the predicted
accuracy (predicted probability of correctly solving a puzzle). Error bars
represent 95\% conﬁdence intervals after Bonferroni correction of the p
value.}
\label{fig:complexityJointAccuracy} 
\end{figure}

Again, we then explored the extent to which cyclomatic complexity moderated
accuracy for puzzles with blindspots versus ones without blindspots for
Python and Java. Again, we applied Bonferroni correction for the $p$-value
threshold to determine statistical significance ($p < 0.025$). The three-way
interaction between presence of blindspot, cyclomatic complexity, and
programming language was significant (Wald $\chi^2(2) = 64.40$, $p < 0.001$).
Figure~\ref{fig:complexityJointAccuracy} shows that, while in Java, the
difference in accuracy for puzzles with and without blindspots increased with
the cyclomatic complexity, in Python, this accuracy difference decreased with
cyclomatic complexity. This significant difference in effect of cyclomatic
complexity on the developers accuracy is surprising. Possible explanations
include differences in the underlying languages, e.g., Python may instill
additional false confidence in the developers, whereas Java's verbosity has
the effect of discouraging developers from making rash decisions. This
hypothesis is supported by our finding that developers rated Java puzzles as
more complex than Python puzzles (see Section~\ref{sec:RQ2and3}). Overall,
these results suggest further study is warranted to understand the different
ways in which code of differing cyclomatic complexity may affect programs in
different languages, and whether such differences can be used to reduce the
pitfalls of developers creating security vulnerabilities as a result of API
blindspots.

\begin{tcolorbox} 
(RQ1, based on cyclomatic complexity) For Python, developers were more likely
to correctly solve puzzles without blindspots than puzzles with blindspots
for low-complexity (but not medium- and high-complexity) puzzles. For Java,
the opposite was true, with developers more likely to correctly solve puzzles
without blindspots than puzzles with blindspots for high-complexity (but not
medium- or low-complexity) puzzles.
\end{tcolorbox}

\subsection{Blindspots' Effect on Difficulty, Clarity, Familiarity, and Confidence}
\label{sec:RQ2and3}

Next, we were interested in learning if the way developers perceive the
puzzles correlates with the developers' ability to solve the puzzles
correctly. For RQ2, we considered the developers perception of the puzzles as
more difficult, less clear, and less familiar. For RQ3, we considered the
developers' confidence in their solutions.

We used a set of multilevel regression models to accommodate for the
hierarchical data structure (recall that each set of six puzzles (level 1
data) is nested within each developer (level 2 data)). The dependent
variables of the four models were the numerical ratings of difficulty,
clarity, familiarity, and confidence, respectively. In each model, we
considered the random effect of the intercept to accommodate for
inter-individual differences in overall ratings of the respective dimension.

For Python, we found no significant effects of presence of a blindspot in a
puzzle on the four dependent variables (all $p > 0.05$), suggesting that
developers' perception of the puzzles did not differ as a function of
presence of a blindspot in the puzzle. The Bayesian statistical analyses
showed that all non-significant effects supported the acceptance of the null
hypothesis (all Bayes Factors $< 0.0007$).

In the Java-puzzle study~\cite{Oliveira18soups}, the developers' perceptions
did not differ as a function of the presence of blindspot in puzzles (all $p
> 0.05$).

We then, again, combined the Java and Python datasets and treated the
language as a moderator in the models. We
found a significant effect of programming language on difficulty (Wald
$\chi^2(1) = 22.11$, $p < 0.001$). That is, developers rated Java puzzles
overall as more difficult than Python puzzles, providing some support for our
earlier hypothesis that some Python puzzles may seem so simple to the
developers that they are even more likely to overlook blindspots in APIs than
they are in more complex puzzles. All other main and interaction effects in
these models (one for each of difficulty, clarity, familiarity, and
confidence) were not significant (all $p > 0.10$). The Bayesian statistical
analyses showed that the Bayes Factor of the non-significant effect for confidence was
0.94, suggesting insensitivity of the present data to detect this effect,
meaning that we cannot draw a conclusion for these factors. In contrast, the
Bayes Factors of the non-significant effects for clarity and familiarity supported the
acceptance of the null hypothesis (all Bayes Factors $< 0.01$).

\begin{tcolorbox}
(RQ2 and RQ3) 
We find that whether the developers considered puzzles to be more difficult,
less clear, and less familiar did not have an effect on their ability to
correctly solve the puzzles. Our statistical analysis supports that
developers rated Java puzzles as more difficult than Python puzzles. We find
that puzzle clarity and familiarity had no effect on the developers' ability
to solve the puzzles, whereas our data was insufficient to draw a conclusion
with respect to the develoers' confidence in their answer.
\end{tcolorbox}

\subsection{Developers' Traits Effect on Ability to Solve Puzzles}
\label{sec:RQ4and5and6}

We next set out to determine if developers' traits affected their ability to
correctly solve puzzles that use APIs with blindspots. For RQ4, we considered
the developers' higher cognitive functions, such as reasoning, working
memory, and processing speed. For RQ5, we considered the developers'
professional experience and expertise. For RQ6, we considered the developers'
levels of conscientiousness and openness and levels of neuroticism and
agreeableness.

For this analysis, we constructed an ordinal outcome variable with a range of
0 to 4 consisting of the number of puzzles with blindpots that a developer
solved correctly. (Recall that each developer attempted to solve 6 puzzles, 4
of which used APIs with blindspots.) We conducted ordinal logistic
regressions using this ordinal outcome variable to test the effect of each
trait a developer possesses on their ability to solve puzzles with blindspots.

We conducted a separate model for each of the three research questions. For
RQ4, measuring cognitive functions, the independent variables consisted of
the Oral Symbol Digit Test from the NIH toolbox and the backward counting,
backward digit span, category fluency, immediate word list recall, delayed
word list recall, and number series from the Brief Test of Adult Cognition by
Telephone (BTACT)~\cite{tun2006}.
Immediate word list recall is a measure of short-term memory, while delayed
word recall is a measure of one's long-term memory. The Oral Symbol Digit
Test is a measure of one's processing speed. The backward counting is a
measure of immediate and delayed episodic memory, the backward digit span is
a measure of working memory span, and the category fluency test measures
verbal fluency, executive functioning, and speed of
processing~\cite{Lachman14}. We measured inductive reasoning using a number
series completion task~\cite{Salthouse87, Schaie96}.

Among all cognitive measures, only the delayed word list recall task showed a
significant effect ($B = 0.20$, $z = 2.37$, $p = 0.02$, odds ratio $= 1.23$).
Developers with better long-term memory were more likely to correctly solve
puzzles with blindspots. None of the other predictors showed a significant
effect (for each, $p > 0.10$). The Bayesian statistical analyses showed that
all non-significant effects supported the acceptance of the null hypothesis
(all Bayes Factors $< 0.002$), meaning that the data support the conclusion
that these factors do not affect developers' ability to correctly solve
puzzles.

In the Java-puzzle study~\cite{Oliveira18soups}, the dataset was limited for
cognitive measures due to missing data, and there were no significant effects
observed for any of the three cognitive measures on blindspot puzzle accuracy
(all $p > 0.05$).

When comparing the two programming languages directly regarding the impact of
cognitive measures on the programmers' ability to solving puzzles, we
examined each cognitive measure individually (i.e., in a separate model). In
the original study~\cite{Oliveira2014}, a technical error led 35 Java
developers to have missing data on some of the cognitive measures. For this
analysis, none of the cognitive measures showed a significant interaction
with programming language (all $p > 0.10$), suggesting that the null effect
of cognitive abilities on solving blindspot puzzles did not vary as a
function of programming language. The Bayesian statistical analyses showed
that all non-significant interactions supported the acceptance of null
hypothesis (all Bayes Factors $< 5 \times 10^{-6}$), meaning that findings are
consistent across Java and Python, and the  programming
language of the puzzles is unlikely to have an effect on which cognitive
functions affect developers' ability to solve puzzles.

\begin{tcolorbox}
(RQ4) We conclude that developers with better long-term memory recall are
more likely to correctly solve puzzles with blindspots. Short-term memory,
processing speed, episodic memory, and memory span do not affect developer's
ability to correctly solve puzzles with blindspots. The programming language
of the puzzles did not affect which cognitive measures has an effect on the
developers ability to solve puzzles correctly.
\end{tcolorbox}

For RQ5, measuring professional experience and expertise, the independent
variables consisted of years of programming, technical proficiency score, and
Python and Java skills. Four developers had missing data on (some of) the
technical expertise or experience measures, so we excluded them from our
analysis in this model, so this model was based on 125 samples. The Java
analysis was based on the full 109 developers' samples~\cite{Oliveira18soups}.

For Python, none of the three predictors of technical expertise and
experience predicted blindspot puzzle accuracy (for each, $p > 0.10$). The
Bayesian statistical analyses showed that all non-significant interactions
supported the acceptance of the null hypothesis (all Bayes Factors $< 0.0001$).

In the Java-puzzle study~\cite{Oliveira18soups}, none of the three predictors
of experience and expertise predicted blindspot puzzle accuracy (all $p >
0.05$).

When directly comparing the two programming languages regarding the impact of
technical expertise and experience on the programmers' ability to solve
puzzles, none of the technical expertise and experience showed a significant
interaction with programming language (all $p \geq 0.10$), suggesting that the
null effect of technical expertise and experience on solving blindspot
puzzles did not vary as a function of programming language. The Bayesian
statistical analyses showed that all these non-significant effects supported
the acceptance of the null hypothesis (all Bayes Factors $< 0.002$), suggesting that
there is no relationship between professional experience and expertice and
the developers' ability to correctly solve puzzles.

\begin{tcolorbox}
(RQ5) We conclude that, surprisingly, professional experience and expertice
do not improve the developers' ability to solve puzzles with blindspots
correctly. This result was consistent across the two languages.
\end{tcolorbox}

\begin{figure}[t]

\centerline{%
\includegraphics[width=.75\columnwidth]{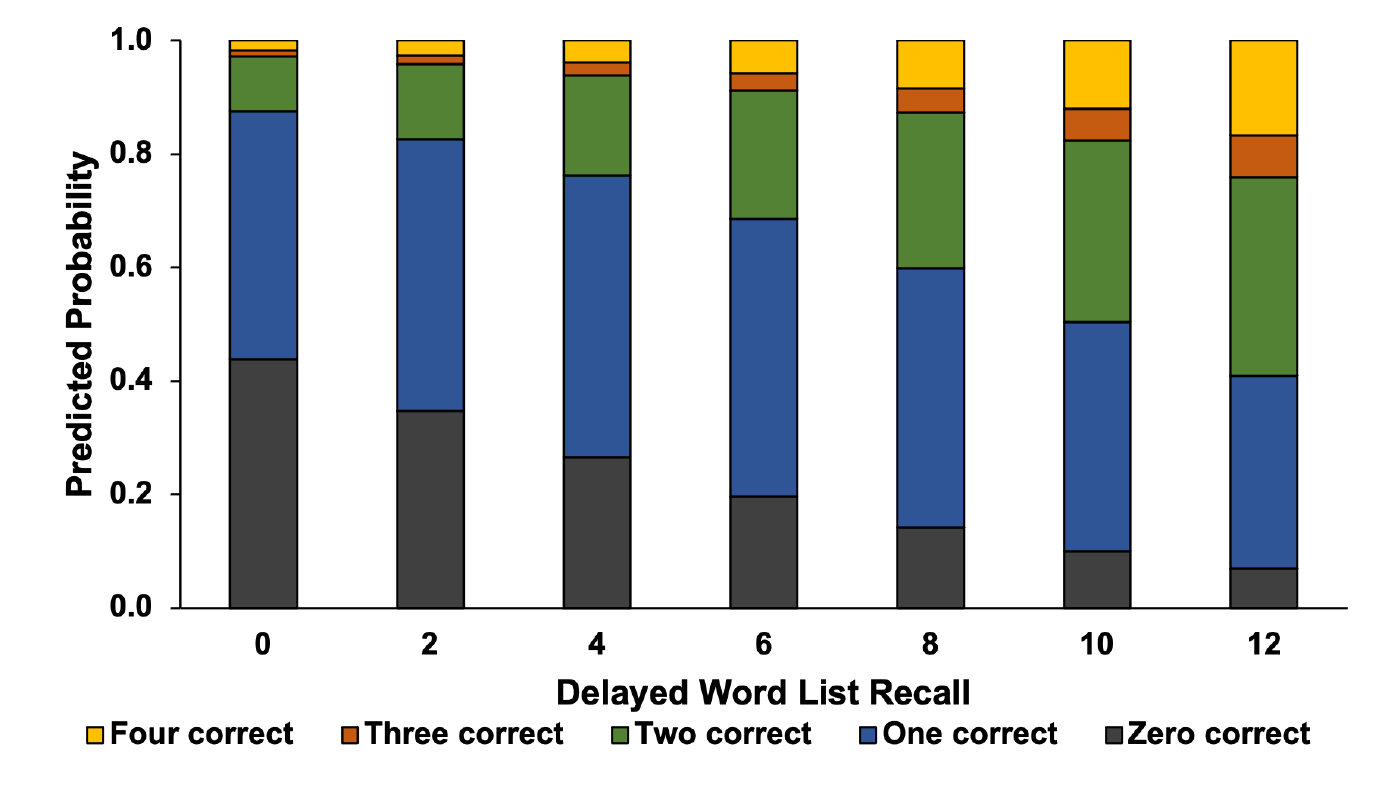}
}

\caption{Effect of long-term memory on puzzle solution accuracy for Python
puzzles with blindspots. The x-axis represents the delayed word list recall
task score (a measure of long-term memory), with higher scores reflecting better
long-term memory. The y-axis shows the predicted probability of correctly
solving a given number (from 0 to 4) of blindspot puzzles; higher accuracy in
solving blindspot puzzles is associated with darker colors.}
\label{fig:delayedWordListRecall} 
\end{figure}

For RQ6, the independent variables consisted of five personality traits:
agreeableness, conscientiousness, extraversion, neuroticism, and openness. We
measured these personality traits use the Big Five Inventory (BFI)
questionnaire~\cite{srivastava99}. The questionnaire is composed of 44
personality statements used to assess the five personality traits by having
the participant rate the level they feel they endorse the given personality
statement using a Likert scale. For this research question (unlike RQ5), our
dataset covered all 129 participants.

For Python, none of the personality traites showed a significant effect (all
$p > 0.18$). The Bayesian statistical analyses showed that all these
non-significant effects supported the acceptance of the null hypothesis (all
Bayes Factors $< 1 \times 10^{-10}$), suggesting that there is no
relationship between the personality traits and the developers' ability to
correctly solve puzzles.

In the Java-puzzle study~\cite{Oliveira18soups}, the effect of openness on
blindspot puzzle accuracy was significant ($p < 0.001$). That is, greater
openness as a personality trait in developers was associated with greater
accuracy in solving blindspot puzzles. None of the other personality
dimensions showed significant effects (all $p > 0.05$).

\begin{figure}[t]

\begin{center}
{\small Java} \\
\includegraphics[width=.75\columnwidth]{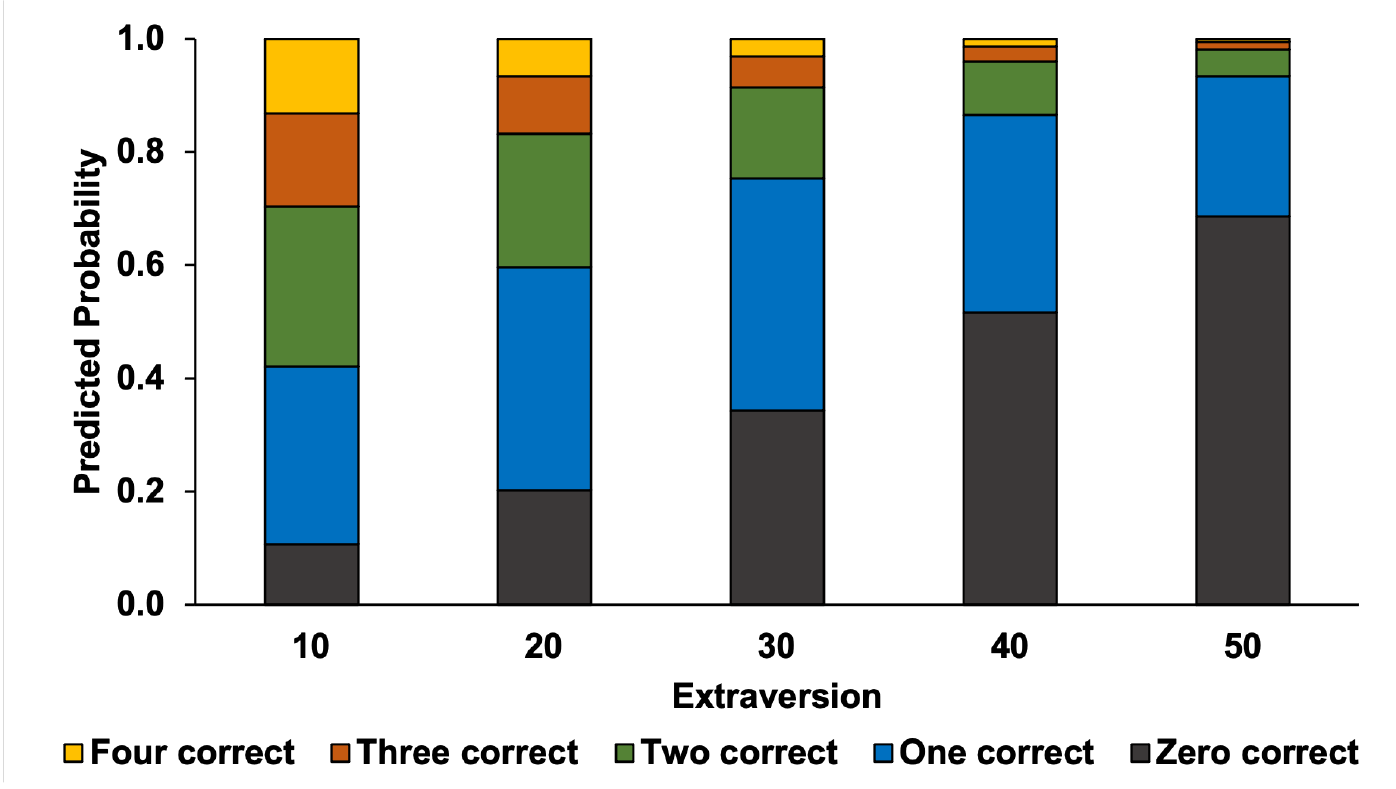} \\
{\small Python} \\
\includegraphics[width=.75\columnwidth]{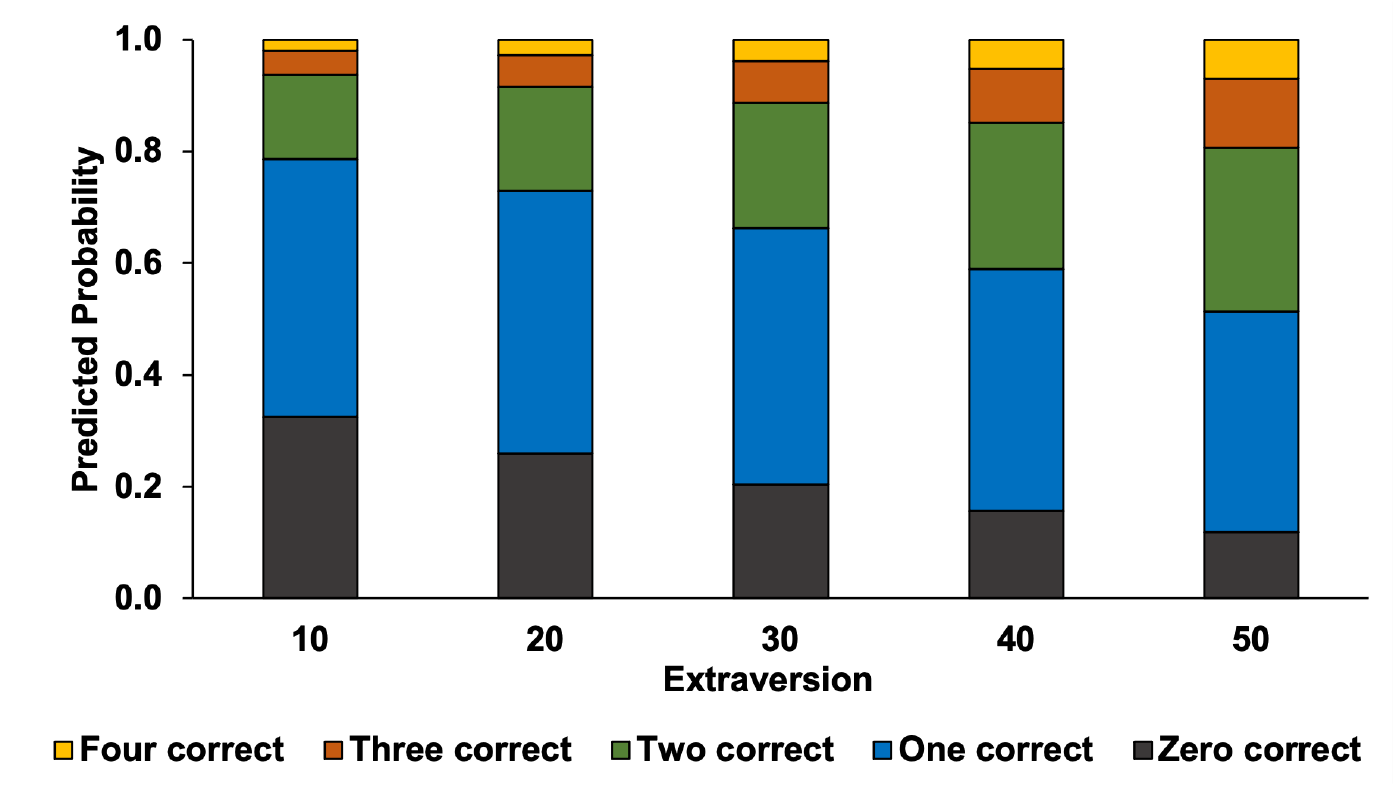}
\end{center}

\caption{Association between extraversion and accuracy for Java (top) and
Python (bottom) puzzles with blindspots. The x-axis represents the level of
extraversion, with higher scores reflecting more extraversion. The y-axis
shows the predicted probability of correctly solving a given number of
blindspot puzzles (from 0 to 4); higher accuracy in solving blindspot
puzzles is associated with darker colors.}
\label{fig:extraversion} 
\end{figure}

\begin{figure}[t]

\begin{center}
{\small Java} \\
\includegraphics[width=.75\columnwidth]{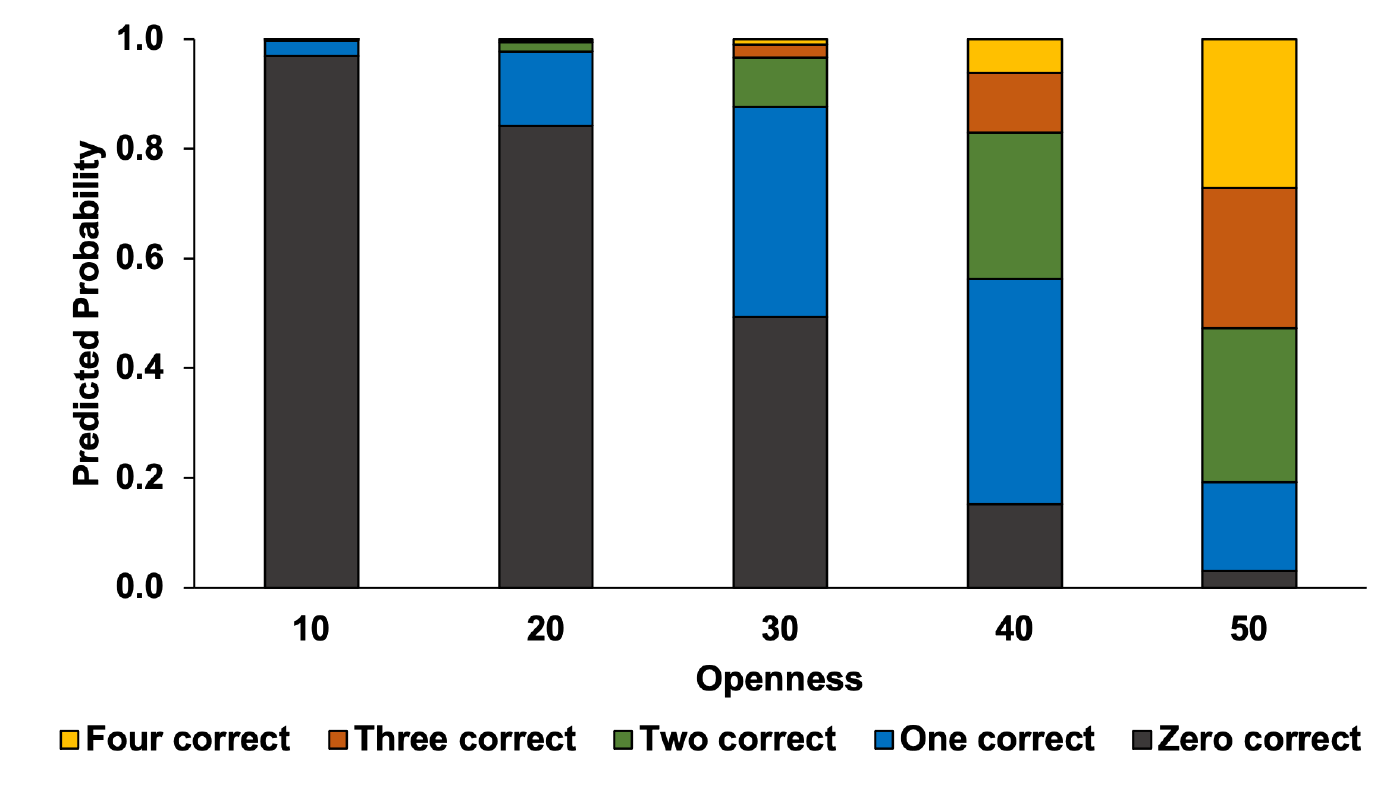}\\
{\small Python} \\
\includegraphics[width=.75\columnwidth]{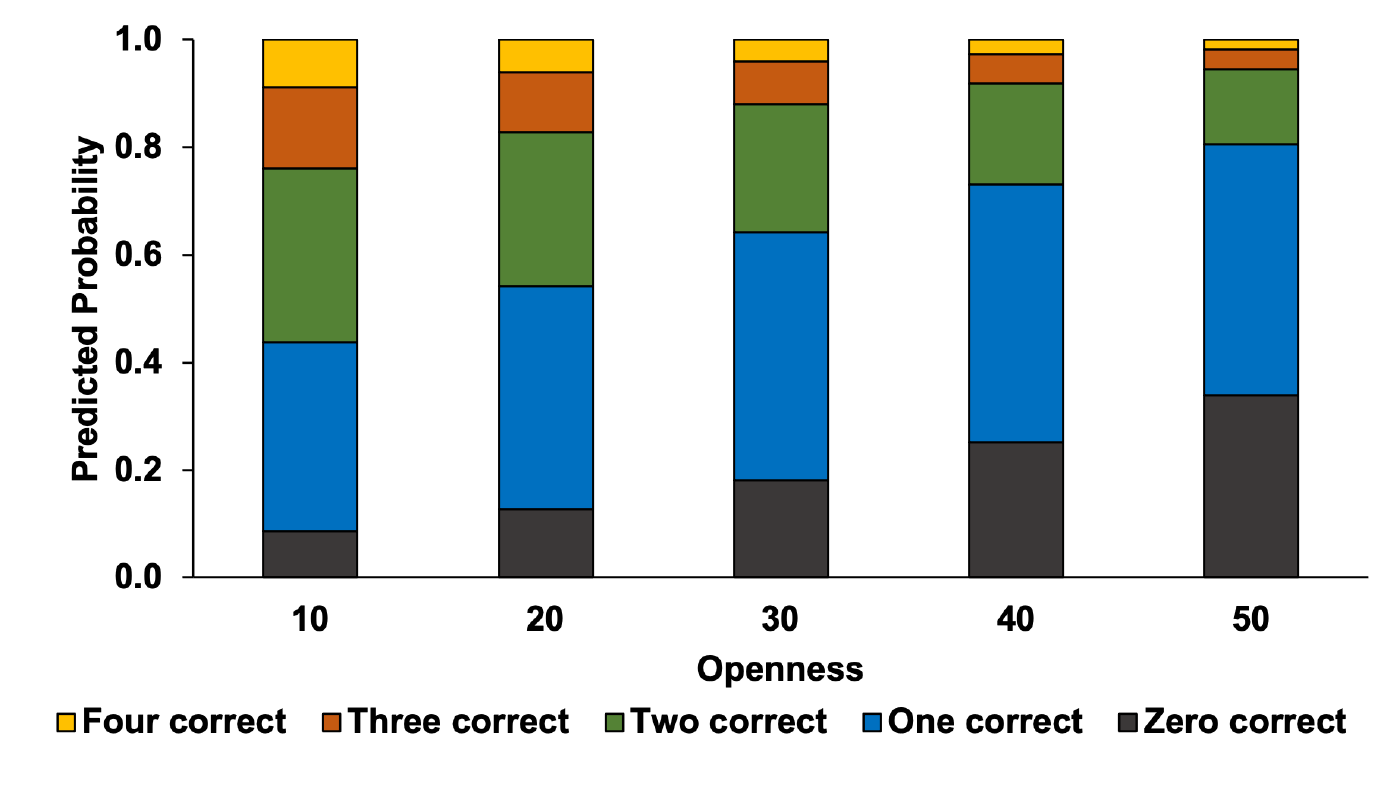}
\end{center}

\caption{Association between openness and accuracy for Java (top) and Python
(bottom) puzzles with blindspots. The x-axis represents the level of openness,
with higher scores reflecting more openness. The y-axis shows the predicted
probability of correctly solving a given number of blindspot puzzles (from 0
to 4); higher accuracy in solving blindspot puzzles is associated with darker
colors.}
\label{fig:openness} 
\end{figure}

When directly comparing the two programming languages regarding the impact of
personality traits on the programmers' ability to solve blindspot puzzles,
the interaction between extraversion and programming language (Wald
$\chi^2(1) = 5.95$, $p = 0.01$) as well as the interaction between openness
and programming language (Wald $\chi^2(1) = 13.50$, $p < 0.001$) were
significant. That is, developers with lower extraversion (see
Figure~\ref{fig:extraversion}) and higher openness (see
Figure~\ref{fig:openness}) were more accurate in solving Java blindspot
puzzles, while these effects did not hold for Python puzzles. The
interactions between agreeableness, conscientiousness, as well as neuroticism
and programming language, however, were not significant (all $p > 0.09$). The
Bayesian statistical analyses showed that all these non-significant
interactions supported the acceptance of the null hypothesis (all Bayes
Factors $< 10^{-5}$).

\begin{tcolorbox}
(RQ6) We conclude that developers with lower extraversion and higher openness
as personality traits were more accurate in solving Java puzzles with
blindspots. In contrast, developers' personality traits were not associated
with their accuracy to solve Python puzzles with blindspots.
\end{tcolorbox}

\section{Discussion, Findings, and Implications}
\label{sec:discussion}

Most importantly, our study confirms that developers are less likely to
correctly understand code that uses APIs with blindspots, even those that
cause well-known vulnerabilities. This effect was most pronounced for APIs
that dealt with I/O for Java; for Python, the effect was seen for all three
API types.

Importantly, developers' expertice and experience did not help mitigate the
risks associated with blindspots. Nor did the developers' perception of
puzzle difficulty, clarity, and their familiarity with the involved concepts
affect their success. Unfortunately, our data was insufficient for us to
determine whether developers' confidence in their solution indicated a higher
chance that their solution was correct. Given the large size of our user
study, the fact that this result was inconclusive suggests that the effect of
confidence may, at least, not be as strong as one might expect.

The only factor about the developers that had a consiste effect on their
ability to correctly solve puzzles was their long-term memory. Developers
with better long-term memory were more likely to correctly solve puzzles.

Interestingly, we observed opposite results for Python and Java with respect
to puzzle complexity. For Python, developers were far more often correct when
solving low-complexity puzzles with APIs without blindspots than with
blindspots, but there was virtually no difference for high-complexity. For
Java, the opposite was true, with developers far more often correct when
solving high-complexity puzzles with APIs without blindspots than with
blindspots. We speculate that developers may have a false sense of security
for low-complexity Python puzzles, which causes them more often to be blind
to the blindspots. The developers rated the Java puzzles to be more complex
than the Python puzzles, so it is possible that the false sense of security
does not surface for Java puzzles. Instead, the more expected lower accuracy
rate for more complex puzzles dominates. A related interesting observation is
that for Java, the accuracy for puzzles without blindspots increased with
complexity.

\section{Related Work}
\label{sec:related}

Our study extends our earlier work~\cite{Oliveira18soups} that focused on how
developers reason about Java APIs that have blindspots. This paper develops
22 new puzzles that use Python APIs, 16 of which have known blindspots, and
replicates our earlier study. This replication improves the data collection
methodology that allows for new analyses, allows us to generalize our
findings across languages and explore differences between Python and Java,
and uses Bayesian statistic analysis to make stronger claims than the prior
study. The Java-based study has led to research into understanding why
developers make security mistakes~\cite{Parker20, Votipka20}, gaining insight
into the developers' rationale in making API-use decisions~\cite{Linden20},
and evaluating the usability of security APIs~\cite{Wijayarathna19}. Our
replication study provides further support for that work. 

Cappos et al.~\cite{Cappos2014} proposed that software vulnerabilities are a
blindspot in developers' heuristic-based decision making mental models.
Oliveira et al.~\cite{Oliveira2014} further showed that security is not a
priority in the developers' mindsets while coding; however, that developers
do adopt a security mindset once primed about the topic. Our work complements
and extends previous investigations on the effect of API blindspots on
writing secure code, and in determining the extent to which developers'
characteristics (perceptions, expertise, experience, cognitive function, and
personality) influence such capabilities.

\subsection{API usability}

The study of API usability focuses on how to design APIs in a manner that
reduces the likelihood of developer errors that can create software
vulnerabilities~\cite{Myers2016}. Such research identifies common pitfalls in
API design. For example, a study showed that the very popular factory design
pattern~\cite{GoF1995} is detrimental to API usability because when
incorporated into an API it was difficult to use~\cite{Ellis2007}.

Most studies of API usability have focused on non-security considerations,
such as examining how well programmers can use the functionality that an API
intends to provide. Our work is, thus, a significant departure from this
research direction, although it shares many of the same methodologies.

Two of the few existing studies on security-related API usability were
conducted by Coblenz et al.~\cite{Coblenz2017, Coblenz2016} and by Weber et
al.~\cite{Weber2017}. Stylos and Clarke~\cite{Stylos2007} had concluded that
the immutability feature of a programming language (i.e., complete
restriction on an object to change its state once it is created) was
detrimental to API usability. Since this perspective contradicted the
standard security guidance (\textit{``Mutability, whilst appearing innocuous,
can cause a surprising variety of security problems''}~\cite{javasecurity,
certjava}), Coblenz et al. investigated the impact of immutability on API
usability and security. From a series of empirical studies, they concluded
that immutability had positive effects on both security and
usability~\cite{Coblenz2016}. Based on these findings they designed and
implemented a Java language extension to realize these
benefits~\cite{Coblenz2017}.

Recent work has investigated the usability of cryptographic APIs. Nadi
et al.~\cite{Nadi2016} identified challenges developers face when using Java
Crypto APIs, namely poor documentation, lack of cryptography knowledge by the
developers, and poor API design. Acar et al.~\cite{Acar17SP} conducted
an online study with open source Python developers about the usability of the
Python Crypto API. In this study, developers reported the need for simpler
interfaces and easier-to-consult documentation with secure, easy-to-use code
examples.

In contrast to previous work, our study focuses on understanding blindspots that
developers experience while working with general classes of API functions.

A significant body work has focused on automatically inferring models of API
use or of APIs themselves to document API use practice. This work has spanned
serial systems~\cite{Dallmeier10, Beschastnikh11fse, Ghezzi14,
Beschastnikh13icse, Beschastnikh11tool-demo-fse, Ghezzi14,
Beschastnikh13icse, Beschastnikh15tse, Krka14fse}, distributed
systems~\cite{Beschastnikh14icse, Beschastnikh11osr}, and
resource-constrained systems~\cite{Ohmann14ase}. Such work typically uses
execution traces to capture API use patterns, which can also be used to
finding the location of a defect~\cite{Dallmeier05, Jones02} and the root
cause of a defect~\cite{Johnson20icse}. Visualizing executions of models of
groups of executions can further help locate bugs~\cite{Jones02} and
understand software behavior~\cite{Beschastnikh16cacm, Beschastnikh20tosem}.
In contrast, our work focuses on understanding how developers act when faced
with APIs that contain blindspots, and is complementary to these techniques
that analyze APIs themselves or patterns of how developers use them.

\subsection{Programming language design}

Usability in programming language design has been a long-standing concern.
Stefik and Siebert~\cite{Stefik2013} showed that syntax used in a programming
language was a significant barrier for novices. Research has also empirically
compared programming languages, in particular, for whether some languages
cause developers to create more bugs than others~\cite{Berger19}. Of course,
some languages are designed to make it impossible to make certain kinds of
errors, e.g., Java makes certain memory-use errors impossible by
automatically managing memory use. Our work has the potential to contribute
to programming language design, since our focus is on understanding security
blindspots in API function usage, and the function traits that exacerbate the
problem.

It is common for systems to exhibit emerging properties. Such properties are
not explicitly coded by the developers into the system, but emerge either by
design from the language, deployment frameworks, or APIs used by the
system~\cite{Brun07seams, Brun07efts, Brun11icdcs, Brun12cloud, Brun13tdsc},
or without intention from interactions of the systems's componentes. These
emerging properties, particularly the latter kind, may surprise developers,
and could result in blindspots. We have adressed blindspots focusing on a
single API, but future work should similarly explore blindspots caused by
potential component interactions.

\subsection{Developer practices and perceptions of security and privacy}

Balebako et al.\ discussed the relationship between the security and privacy
mindsets of mobile app developers and company characteristics (e.g., company
size, having a Chief Privacy Officer, etc.). They found that developers tend
to prioritize security tools over privacy policies, mostly because of the
language of privacy policies is so obscure~\cite{balebako2014}.

Xie et al.~\cite{xie2011} conducted interviews with professional developers
to understand secure coding practices. They reported a disconnect between
developers' conceptual understanding of security and their attitudes
regarding personal responsibility and practices for software security.
Developers also often hold a ``not-my-problem'' attitude when it comes to
securing the software they are developing; that is, they appear to rely on
other processes, people, or organizations to handle software security.

Witschey et al.~\cite{Witschey2015} conducted a survey with professional
developers to understand factors contributing to the adoption of security
tools. They found that peer effects and the frequency of interaction with
security experts were more important than security education, office policy,
easy-to-use tools, personal inquisitiveness, and better job performance to
promote security tool adoption.

Acar et al.~\cite{Acar2016} and Green and Smith~\cite{Green2016} suggest a
research agenda to achieve usable security for developers. They proposed
several research questions to elicit developers' attitudes, needs, and
priorities in the area of security. Oltrogge et al.~\cite{Oltrogge2015} asked
for developers' feedback on TLS certificate pinning strategy in non-browser
based mobile applications. They found a wide conceptual gap about pinning and
its proper implementation in software due to API complexity.

A survey conducted by Acar et al.~\cite{Acar16SP} with 295 app developers
concluded that developers learned security through web search and peers. The
authors also conducted an experiment with over 50 Android developers to
evaluate the effectiveness of different strategies to learn about app
security. Programmers who used digital books achieved better security than
those who used web searches. Recent research corroborates this finding by
showing that the use of code-snippets from online developer fora (e.g., Stack
Overflow) can lead to software vulnerabilities~\cite{Acar17, Fischer2017,
Unruh2017}.

Certain aspects of software systems, such as security or
fairness~\cite{Dwork12, Galhotra17fse, Brun18fse-nier} are not only difficult
for developers to reason about, it can be difficult for end-users to
understand and describe the relevant requirements~\cite{Grgic-Hlaca18}.
Recent work has aimed to develop APIs for components that can provide
guarantees, e.g., that the component will not exhibit racist or sexist
behavior when applied to future inputs~\cite{Thomas19science,
Metevier19neurips, Agarwal18}. While early work has looked at whether such
APIs help or hurt developers' and data scientists' work on improving system
fairness~\cite{Johnson20fairkit}, fully understanding the implications of
such APIs remains an open problem. Systems that
automatically~\cite{Galhotra17fse, Angell18demo-fse} and
manually~\cite{Tramer17} test systems for these properties are likely to help
developers reason about their uses of these APIs.

Recent studies have investigated the need and type of interventions required
for developers to adopt secure software development practices. Xie et
al.~\cite{Xie2012} found that developers needed to be motivated to fix
software bugs. There has also been some work on how to create this motivation
and encourage use of security tools. Several surveys identified the
importance of social proof for developers' adoption of security
tools~\cite{Murphy2015, Witschey2014, Xiao2014}. Meanwhile the way that tools
communicate with developers has been shown to be a critical aspect of
successful adoption~\cite{Johnson16}.

Research on the effects of external software security consultancy
suggests~\cite{Poller2016} that a single time-limited involvement of
developers with security awareness programs is generally ineffective in the
long-term. Poller et al.~\cite{Poller2017} explored the effect of
organizational practices and priorities on the adoption of developers' secure
programming. They found that security vulnerability patching is done as a
stand-alone procedure, rather than being part of product feature development.
In an interview-based study by Votipka et al.~\cite{Mazurek2018} with a group
of 25 white-hat hackers and software testers on bug finding related issues,
hackers were more adept and efficient in finding software vulnerabilities
than testers, but they had more difficulty in communicating such issues to
developers because of a lack of shared vocabulary.

Most industrial and open-source development happens collaboratively, which
can lead to other pitfalls, such as collaborative conflicts. Tools that
improve developer-awareness can help avoid such pitfalls~\cite{Brun11fse,
Brun13tse, Brun11tool-demo-fse, Sarma03, Sarma12}. We envision similar tools
can be built to warn developers of blindspots in the APIs they're using,
helping them avoid introducing vulnerabilities into their code.

Our work on studying how developers use APIs and how blindspots affect them,
and their code, complements these studies that have not focused on
blindspots. Together, this work is building a better understanding of the
developers' processes, and how the tools and APIs at their disposal improve,
or stand in their way, of writing high-quality, secure programs.

\section{Contributions}
\label{sec:contributions}

We have replicated our earlier, Java-based controlled experiment studying the
effect of APIs with blindspots on developers~\cite{Oliveira18soups}. Our
replication applies to Python and involves 129 new developers and 22 new
APIs. We found that APIs with blindspots statistically significantly reduce
the developers' ability to correctly reason about the APIs in both languages,
but that the effect is more pronounced for Python. Professional experience
and expertice failed to minitage this reduction, with long-term professionals
with many years of experience making mistakes as often as relative novices.
These findings suggest that blindspots in APIs are a serious problem across
languages, and that experience and education alone are insufficient to
overcome it. Tools are needed to help developers recognize blindspots in APIs
as they write code that uses those APIs, warning the developers and reducing
the risk of the introduction of vulnerabilities.

Interestingly, for Java, the ability to correctly reason about APIs with
blindspots improved with complexity of the code, whereas for Python, the
opposite was true. We hypothesize that Python developers are less likely to
notice potential for vulnerabilities in complex code than in simple code,
whereas Java developers are more likely to recognize the extra complexity and
apply more care, but are more careless with simple code. This finding
suggests that, while blindspots likely have negative effects across
programming languges, there are important differences among languages,
warranting both further studies of more languges with respect to blindspots,
and deeper studies into whether the blindspots' unique effects on each
language can be used to reduce the potentially resulting vulnerabilities.

\section{Acknowledgments}
This work is supported by the National Science Foundation under grants
CCF-1453474, 
CNS-1513055, 
CNS-1513457, 
CNS-1513572, 
and CCF-1564162. 
We wish to thank Daniela Oliveira, Muhammad Sajidur Rahman, Rad Akefirad,
Donovan Ellis, Eliany Perez, Rahul Bobhate, Lois A. DeLong, and Justin Cappos
for their contributions to data collection, puzzle creation, and the
Java-based study.

\balance

\bibliographystyle{plain}
\bibliography{biblio}

\end{document}